\documentclass[aps,twocolumn,prex,floatfix,longbibliography,superscriptaddress,nofootinbib,pra]{revtex4-2}
\pdfoutput=1

\usepackage[unicode]{hyperref}
\usepackage{physics}
\usepackage[export]{adjustbox}
\usepackage{enumitem}
\usepackage{amsmath}

\hypersetup{
    unicode=true, 
    plainpages=false,
    colorlinks=true,
    linkcolor=blue,
    citecolor=blue,
    filecolor=blue,
    urlcolor=blue
}
\usepackage{xcolor}
\usepackage{amsmath}
\usepackage[utf8]{inputenc} 
\usepackage[T1]{fontenc}    
\usepackage{url}            
\usepackage{booktabs}       
\usepackage{amsfonts}       
\usepackage{nicefrac}       
\usepackage{microtype}      
\usepackage{lipsum}
\usepackage{graphicx}
\graphicspath{{media/}}     
\usepackage{braket}
\usepackage[normalem]{ulem} 

\begin{document}

\title{Frictional work and entropy production in integrable and non-integrable spin chains}

\author{Vishnu Muraleedharan Sajitha}
\email{v.muraleedharansajitha@uq.edu.au}
\affiliation{University of Queensland -- IIT Delhi Academy of Research, Hauz Khas, New Delhi 110016, India}
\affiliation{ARC Centre of Excellence for Engineered Quantum Systems, School of Mathematics and Physics, University of Queensland, St Lucia, Queensland 4072, Australia}
\affiliation{Department of Physics, Indian Institute of Technology, Delhi, New Delhi 110016, India}

\author{Matthew J. Davis}
\email{mdavis@uq.edu.au}
\affiliation{ARC Centre of Excellence for Engineered Quantum Systems, School of Mathematics and Physics, University of Queensland, St Lucia, Queensland 4072, Australia}
\author{L. A. Williamson}
\email{lewis.williamson@uq.edu.au}
\affiliation{ARC Centre of Excellence for Engineered Quantum Systems, School of Mathematics and Physics, University of Queensland, St Lucia, Queensland 4072, Australia}

\date{\today}

\begin{abstract}
The maximum work extractable from a quantum system is achieved when the system is driven adiabatically. Frictional work $\langle W\rangle_\mathrm{fric}$ then quantifies the difference in work output between adiabatic and non-adiabatic driving. Here we show that frictional work in a non-integrable spin chain is well-characterized by the diagonal entropy production $\Delta S_\mathrm{d}$ associated with the build up of quantum coherence. We show that, over a broad range of parameters, $\langle W\rangle_\mathrm{fric}\approx T_\tau\Delta S_\mathrm{d}$, with $T_\tau$ the effective temperature of the final time-evolved state. The relationship breaks down for fast protocols at low temperatures, in which case frictional work is instead well-described by the quantum relative entropy between the time-evolved state and a Gibbs-state approximation of the adiabatic state. We compare our results to those obtained from an integrable spin chain, in which case the system is no longer described by a single temperature. In this case, the frictional work is described by a sum of terms for each independent subspace of the spin chain, which are at different effective temperatures. Finally, we show how integrability breaking can enhance work extraction in the adiabatic limit, but degrade work extraction in sufficiently non-adiabatic regimes.
\end{abstract}

\maketitle

\section{Introduction}

Understanding the role of coherence in work extraction is a fundamental point of interest in the field of quantum thermodynamics~\cite{vinjanampathy2016}. Although carefully constructed protocols have utilized coherence in thermodynamic tasks~\cite{uzdin2015, klatzow2019, korzekwa2016, kammerlander2016, francica2020, williamson2024, uzdin2016, scully2003, PhysRevE.109.014102}, in many cases the build up of coherence is detrimental to work output~\cite{horodecki2013}. In particular, in non-adiabatic work extraction, coherences can build up in the energy eigenbasis, which are then lost under a projective energy measurement --- a type of ``quantum friction''~\cite{kosloff2002,feldmann2003}. The understanding and mitigation of these frictional effects is important, for example, for achieving high efficiency and power output in quantum heat engines~\cite{feldmann2006,rezek2006,salamon2009,Alecce_2015,kosloff2013}.

\begin{figure}[h]
    \centering
    \includegraphics[width=\linewidth]{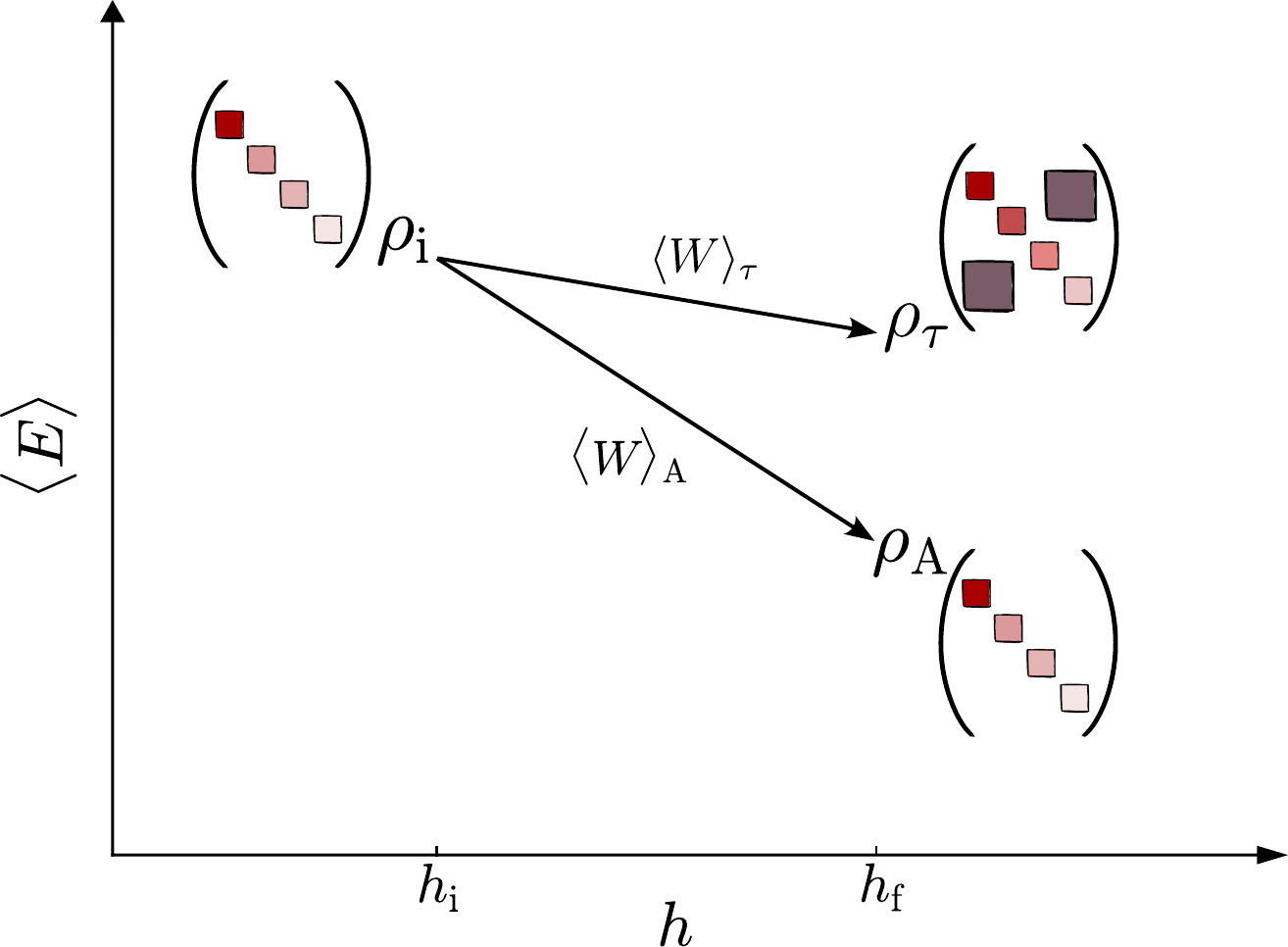}
\caption{Schematic of work extraction in finite-time and adiabatic protocols. The system begins in a thermal state $\rho_\mathrm{i}$, which is diagonal in the initial energy eigenbasis. Work is extracted by tuning $h$. Non-adiabatic extraction (duration $\tau$) results in coherences in the final state $\rho_\tau$ with respect to the final energy basis. In the adiabatic limit, the final state $\rho_\mathrm{A}$ remains diagonal. The difference in work output between the non-adiabatic and adiabatic processes defines the frictional work $\langle W\rangle_\mathrm{fric}\equiv\langle W\rangle_\tau-\langle W\rangle_\mathrm{A}$.}   
    \label{fig:protocol}
\end{figure}

The minimal work principle states that, for an isolated system initially in thermal equilibrium, the most work is extracted (or least work is expended) when the system is driven adiabatically, $\langle W\rangle_\tau\ge\langle W\rangle_\mathrm{A}$~\cite{PhysRevE.71.046107}. Here $\langle W \rangle_\tau\equiv\operatorname{Tr}[\rho_\tau\hat{H}_\mathrm{f}]-\operatorname{Tr}[\rho_\mathrm{i}\hat{H}_\mathrm{i}]$ is the work output for a non-adiabatic process of duration $\tau$, {with $\langle W\rangle_\tau<0$ corresponding to energy out of the system, and $\langle W\rangle_\mathrm{A}$ is the adiabatic ($\tau\rightarrow\infty$) work output. We define work via the two-point projective measurement scheme~\cite{PhysRevE.75.050102,PhysRevX.5.031038}, with $\rho_\mathrm{i}$ the initial thermal state, $\rho_\tau$ the time-evolved state, and $\hat{H}_\mathrm{i}$($\hat{H}_\mathrm{f}$) the initial(final) Hamiltonian.} The difference between the adiabatic and non-adiabatic work output is termed \emph{frictional work} (see Fig.~\ref{fig:protocol})~\cite{PhysRevLett.113.260601}, 
\begin{equation}\label{wfric}
\langle W\rangle_{\mathrm{fric}} \equiv \langle W \rangle_\tau - \langle W\rangle_\mathrm{A}\ge 0,
\end{equation}
which quantifies the energy ``wasted'' by using a non-adiabatic drive.

For non-adiabatic driving, the final state $\rho_\tau$ will typically contain coherences with respect to the final energy eigenbasis, which are responsible for frictional effects~\cite{feldmann2012,kosloff2013}. The ensemble of final projective energy measurements then involves a loss of information~\cite{balian1989}. The associated entropy production is quantified by the diagonal entropy \cite{POLKOVNIKOV2011486,PhysRevLett.107.040601,IKEDA2015338,e19040136,kosloff2013},
\begin{equation}\label{eq:Sd}
    S_\mathrm{d}\left(\rho_\tau\right)\equiv S\left(\rho_\tau^\mathrm{diag}\right)
\end{equation}
where
\begin{equation}
    \rho_\tau^\mathrm{diag} \equiv \sum_n \langle n_\mathrm{f} | \rho_\tau|n_\mathrm{f}\rangle |n_\mathrm{f}\rangle \langle n_\mathrm{f}| ,
\end{equation}
is the projection of $\rho_\tau$ onto the final energy eigenbasis $\ket{n_\mathrm{f}}$ and $S(\rho) = -\mathrm{Tr}[\rho \ln \rho]$ is the von Neumann entropy. In general, it can be shown that $S_\mathrm{d}$ increases under unitary driving~\cite{POLKOVNIKOV2011486} and therefore $S_\mathrm{d}$ characterizes the entropy increase intuitively expected from finite-time processes, in contrast to the (conserved) von Neumann entropy.

Plastina and coauthors~\cite{PhysRevLett.113.260601,PhysRevE.99.042105} have established a connection between coherences and frictional work {when the adiabatically evolved state $\rho_\mathrm{A}$} is a Gibbs state with temperature $T_\mathrm{A}$, $\rho_\mathrm{A}=\rho_{T_\mathrm{A}}^\mathrm{therm}$. (Here and below we define $\rho_T^\mathrm{therm}\equiv e^{-\hat{H}_\mathrm{f}/T}/\operatorname{Tr}[e^{-\hat{H}_\mathrm{f}/T}]$ for some specified temperature $T$. We use units where Boltzmann's constant $k_\mathrm{B}\equiv 1$ so that the von Neumann entropy of a Gibbs state coincides with the thermodynamic entropy.) The frictional work then satisfies~\cite{PhysRevLett.113.260601},
\begin{equation}\label{plastina}
 \langle W \rangle_{\mathrm{fric}} = 
 T_\mathrm{A} D(\rho_\tau\|\rho_\mathrm{A})
 \quad (\rho_\mathrm{A} = \rho_{T_\mathrm{A}}^\mathrm{therm}),
\end{equation}
with
\begin{equation}
D(\rho\|\sigma)\equiv\operatorname{Tr}[\rho\log\rho]-\operatorname{Tr}[\rho\log\sigma]
\end{equation}
the quantum relative entropy. Furthermore, as shown in~\cite{PhysRevE.99.042105}, the general identity
\begin{equation}\label{eq:Ddecom}
    D(\rho_\tau\|\rho_\mathrm{A})=\Delta S_\mathrm{d}+D(\rho_\tau^\mathrm{diag}\|\rho_\mathrm{A})
\end{equation}
decomposes the frictional work Eq.~\eqref{plastina} into a population term $T_\mathrm{A} D(\rho_\tau^\mathrm{diag}\|\rho_\mathrm{A})$ and a term arising from coherence $T_\mathrm{A}\Delta S_\mathrm{d}$,
with
\begin{equation}\label{Sdef}
\Delta S_\mathrm{d}\equiv S_\mathrm{d}(\rho_\tau) - S_\mathrm{d}(\rho_\mathrm{A})
\end{equation}
the diagonal entropy production during the finite-time protocol. Noting that $S_\mathrm{d}(\rho_\mathrm{A})=S(\rho)$, this is also the relative entropy of coherence with respect to the final energy eigenbasis, which is a coherence monotone~\cite{PhysRevLett.113.140401,RevModPhys.89.041003}. For slow work processes, the population term can be shown to be small [$D(\rho_\tau^\mathrm{diag}\|\rho_\mathrm{A})=O(\tau^{-2})$]~\cite{PhysRevE.99.042105}, in which case frictional work corresponds predominantly to the build up of coherence,
\begin{equation}\label{eq:Wfricdecom}
    W_\mathrm{fric}\approx T_\mathrm{A}\Delta S_\mathrm{d}\quad (\rho_\mathrm{A} = \rho_{T_\mathrm{A}}^\mathrm{therm}).
\end{equation}

The results Eq.~\eqref{plastina} and Eq.~\eqref{eq:Wfricdecom} as derived in~\cite{PhysRevLett.113.260601,PhysRevE.99.042105} require that $\rho_\mathrm{A}$ be a Gibbs state, which is a substantial restriction. In particular, for $\rho_\mathrm{A}$ to be a Gibbs state, it is necessary that the final energy eigenvalues $E_n^\mathrm{f}$ are related to the initial energy eigenvalues $E_n^\mathrm{i}$ by a constant scaling factor $\lambda$, i.e., $E_n^\mathrm{f}=\lambda E_n^\mathrm{i}$. The temperature $T_\mathrm{A}$ of $\rho_\mathrm{A}$ is then related to the initial temperature $T_\mathrm{i}$ via $T_\mathrm{A}=\lambda T_\mathrm{i}$. This is satisfied in simple systems, for example, a two-level system or a harmonic oscillator. However, in most systems, in particular in many-body interacting systems, 
energy levels will change disproportionately and $\rho_\mathrm{A}$ will deviate from a Gibbs state. This motivates an understanding of $\langle W\rangle_\mathrm{fric}$ and how it relates to coherence that is applicable more generally. Notably, non-equilibrium states of non-integrable quantum systems often exhibit thermal properties according to the eigenstate thermalisation hypothesis~\cite{dalessio2016,mori2018}, with reduced fluctuations of adjacent energy level spacings due to level repulsion and spectral rigidity~\cite{haake2018}. In non-integrable systems we might therefore expect relationships similar to Eq.~\eqref{plastina} and Eq.~\eqref{eq:Wfricdecom} even when $\rho_\mathrm{A}$ is not a Gibbs state.

In this paper we extend the results of Plastina \emph{et al}.\ by showing that Eq.~\eqref{eq:Wfricdecom}, generalised here to include fast work processes, can also describe a non-integrable system with $\rho_\mathrm{A}\ne \rho_{T_\mathrm{A}}^\mathrm{therm}$. Using a quantum spin chain as a representative non-integrable system, we show that frictional work can be directly related to the build up of quantum coherence, with
\begin{equation}\label{main}
\langle W \rangle_{\mathrm{fric}} \approx T_\tau\Delta S_\mathrm{d}.
\end{equation}
The temperature $T_\tau$ is the effective temperature of the time-evolved state $\rho_\tau$, defined as the temperature of the thermal state with the same mean energy, $\operatorname{Tr}[\hat{H}_\mathrm{f}\rho_\tau]=\operatorname{Tr}[\hat{H}_\mathrm{f}\rho_{T_\tau}^\mathrm{therm}]$. For slow work processes we find $T_\mathrm{A}\approx T_\tau$, with $T_\mathrm{A}$ the effective temperature of $\rho_\mathrm{A}$ ($\operatorname{Tr}[\hat{H}_\mathrm{f}\rho_\mathrm{A}]=\operatorname{Tr}[\hat{H}_\mathrm{f}\rho_{T_\mathrm{A}}^\mathrm{therm}]$), and Eq.~\eqref{main} reduces to
\begin{equation}\label{mainTA}
\langle W\rangle_\mathrm{fric}\approx T_\mathrm{A}\Delta S_\mathrm{d}.
\end{equation}
The approximation~\eqref{main} breaks down for fast work processes at low temperatures. In this regime, we find that frictional work can instead be well approximated by
\begin{equation}\label{main1}
    \langle W\rangle_\mathrm{fric}\approx T_\mathrm{A}D(\rho_\tau\|\rho_{T_\mathrm{A}}^\mathrm{therm}).
\end{equation}
{Equation~\eqref{main1} reduces to the equality Eq.~\eqref{plastina} when $\rho_{T_\mathrm{A}}^\mathrm{therm}=\rho_\mathrm{A}$.}

We identify the importance of integrability breaking by comparing our results to those of an integrable spin chain. In the integrable spin chain, a single temperature no longer characterizes the temperature of $\rho_\tau$ and deviation from Eq.~\eqref{main} grows extensively as the number of spins is increased. Additionally, we show that integrability breaking can enhance work extraction in the adiabatic limit, but can be detrimental in the non-adiabatic regime.

{This paper is organized as follows. In Sec.~\ref{sec:system} we introduce our system --- a spin chain described by the transverse-field Ising model with a longitudinal field. Our main results are presented in Sec.~\ref{sec:main}, where we demonstrate the validity of Eq.~(9)-(11) in a non-integrable spin chain. We examine frictional work in an integrable spin chain in Sec.~\ref{sec:integrable}, which establishes the importance of integrability breaking for the results Eq.~(9)-(11) to hold. Finally, in Sec.~\ref{non_integrable} we show how integrability breaking can either enhance or degrade work extraction depending on the duration of the work step. We conclude in Sec.~\ref{conclusion}.} For convenience, a summary of definitions of important quantities used throughout the paper is provided in Table~\ref{definitions}.

\begin{table}[h]
\begin{tabular}{|c|c|}
    \hline
    $\rho_\mathrm{i}$ &Initial Gibbs state $e^{-\hat{H}_\mathrm{i}/T_\mathrm{i}}/\operatorname{Tr}[e^{-\hat{H}_\mathrm{i}/T_\mathrm{i}}]$\\
    \hline
    $\rho_\tau$ &Time-evolved density matrix\\
    \hline
    $\rho_\mathrm{A}$ &Adiabatically-evolved density matrix\\
    \hline
    $\rho_\tau^\mathrm{diag}$ &Projection of $\rho_\tau$ onto $\hat{H}_\mathrm{f}$ eigenbasis\\
    \hline
    $\rho_T^\mathrm{therm}$ &Gibbs state $e^{-\hat{H}_\mathrm{f}/T}/\operatorname{Tr}[e^{-\hat{H}_\mathrm{f}/T}]$\\
    \hline
    $T_\tau$ &{Effective temperature of $\rho_\tau$}\\
    \hline
    {$T_\mathrm{A}$} &{Effective temperature of $\rho_\mathrm{A}$}\\
    \hline
    $\langle W\rangle_\tau$ &Finite-time work\\
    \hline
    $\langle W\rangle_\mathrm{A}$ &Adiabatic work\\
    \hline
    $\langle W\rangle_\mathrm{fric}$ &Frictional work\\
    \hline
    $S$ &Von Neumann entropy\\
    \hline
    $S_\mathrm{d}$ &Diagonal entropy\\
    \hline
\end{tabular}
\caption{\label{definitions} Definitions of important quantities.}
\end{table}


\section{System}\label{sec:system}
We consider the transverse-field Ising model with a longitudinal field, described by the time-dependent Hamiltonian ($\hbar\equiv 1$)
\begin{equation}\label{eq:H}
    \hat{H}(t) = -g \sum_{j=1}^{N} \hat{\sigma}_j^x \hat{\sigma}_{j+1}^x - h(t)\sum_{j=1}^N \hat{\sigma}_j^z + L \sum_{j=1}^{N} \hat{\sigma}_j^x.
\end{equation}
Here $\hat{\sigma}^\mu_{j} $ are spin-1/2 Pauli operators acting on site $ j $ ($[\hat{\sigma}_j^\mu,\hat{\sigma}_j^\nu]=2i\epsilon_{\mu\nu\kappa}\hat{\sigma}_j^\kappa$), $ g $ is the Ising interaction strength and $L$ is the longitudinal field strength. The system is driven by a time-dependent transverse field $h(t)$. We employ periodic boundary conditions $\hat{\sigma}_{N+1}^\nu = \hat{\sigma}_1^\nu$. For $L=0$ the spin chain is integrable and Eq.~\eqref{eq:H} can be diagonalized by mapping the system onto $N$ non-interacting fermions~\cite{PFEUTY197079,mbeng2020quantum,Sachdev_2011}. A non-zero $L$ breaks integrability; for the finite-size systems explored here this integrability breaking is most robust for $g\sim h\sim L$~\cite{kim2013,kim2014}. Quantum spin chains such as Eq.~\eqref{eq:H} can be realised in a variety of experimental platforms~\cite{britton2012,RevModPhys.93.025001,Labuhn2016,browaeys2020} and have been studied extensively as a working substance in engine cycles ~\cite{piccitto2022,revathy2020,PhysRevB.109.024310,PhysRevB.109.224309,jthm-7c2j,wang2020,Solfanelli_2023}.

\begin{figure}[h]
    \centering
    \includegraphics[width=\linewidth]{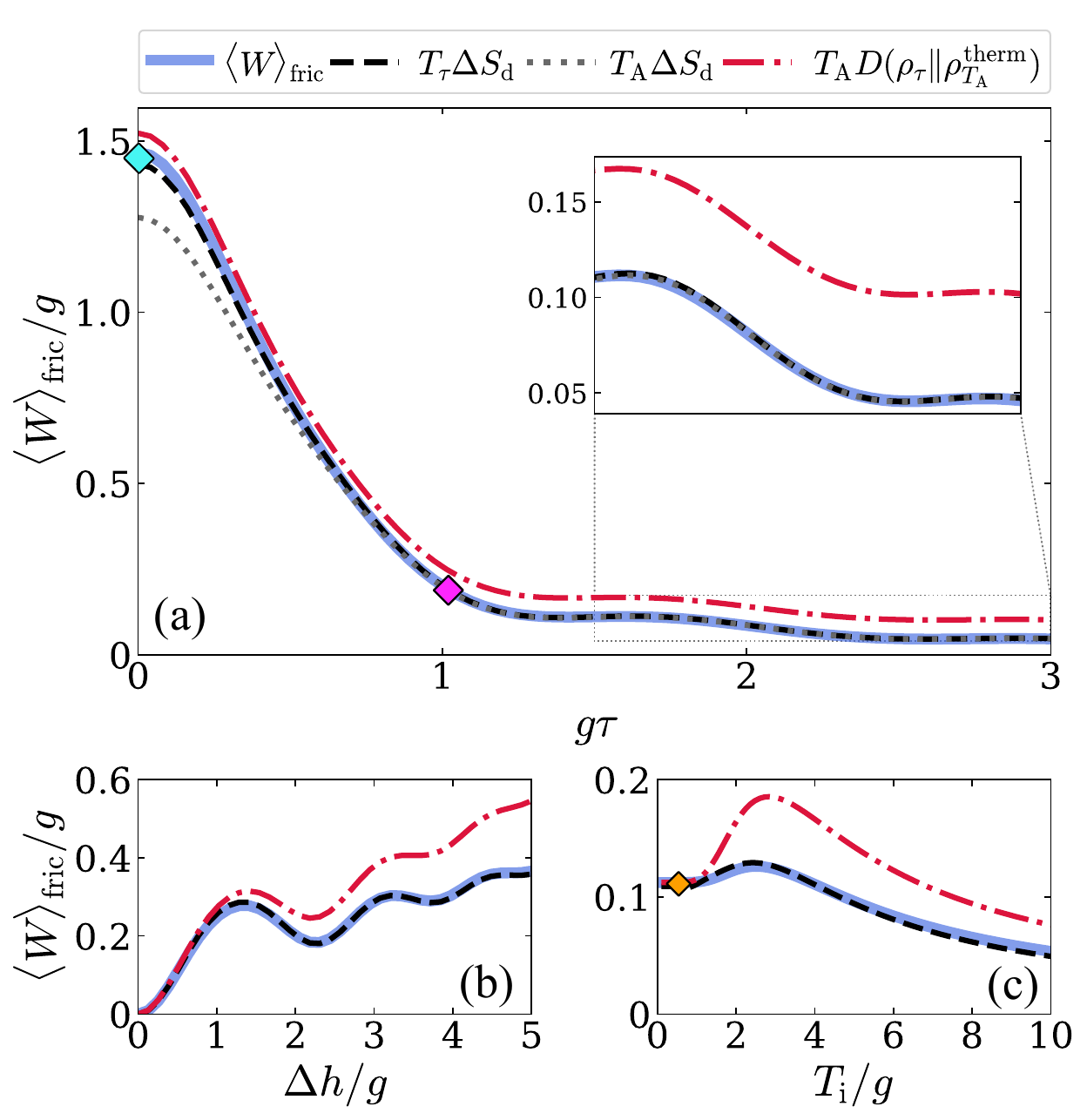}
\caption{Frictional work (solid blue line) is well described by the diagonal entropy production (black dashed line), with a protocol-dependent temperature $T_\tau$ given by the effective temperature of the time-evolved state [Eq.~\eqref{main}]. Agreement is demonstrated for varying (a) protocol durations $\tau$ ($T_\mathrm{i}=4g$, $\Delta h=2g$), (b) work step size $\Delta h$ ($T_\mathrm{i}=4g$, $\tau/\Delta h=0.5g^{-2}$) and (c) initial temperature $T_\mathrm{i}$ ($\tau=1.5g^{-1}$, $\Delta h=2g$). (a) For slow work processes $\tau\gtrsim g^{-1}$, $T_\tau$ is close to the temperature $T_\mathrm{A}$ of the adiabatically evolved state and frictional work is well described by $T_\mathrm{A}\Delta S_\mathrm{d}$ (gray dotted line). The quantum relative entropy relation Eq.~\eqref{main1} is also shown (red dot-dashed line), and typically deviates from $\langle W\rangle_\mathrm{fric}$. Colored diamonds in (a) and (c) mark the parameter regimes for the distributions shown in Fig.~\ref{fig:fig4pop}. All results are for $N=8$, $L=g$ and $h_\mathrm{i}=1.5g$, with $\delta E\approx 4.98g$.}
    \label{fig:fig2new}
\end{figure}

\begin{figure}[h]
    \centering
    \includegraphics[width=\linewidth]{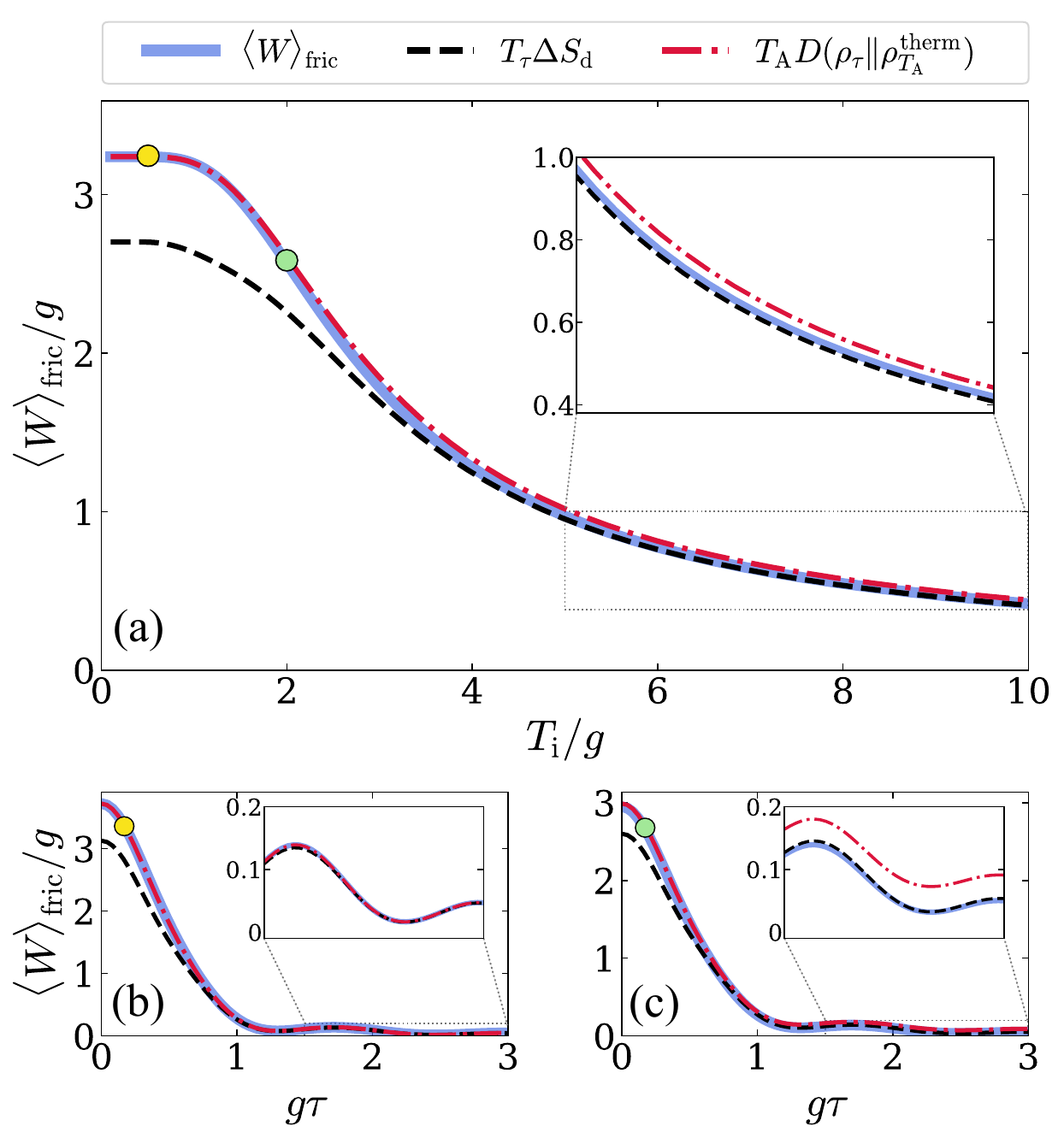}
 \caption{(a) Deviation between frictional work (solid blue line) and $T_\tau\Delta S_\mathrm{d}$ (black dashed line) becomes appreciable at low temperatures for fast work processes (here $\tau=0.2g^{-1}$). In this regime, frictional work is well-described by the quantum relative entropy relation Eq.~\eqref{main1} (red dot-dashed line). (b) and (c) show results as a function of varying $\tau$ for $T_\mathrm{i}=0.5g$ and $T_\mathrm{i}=2g$ respectively, and demonstrate growing deviation between $\langle W\rangle_\mathrm{fric}$ and Eq.~\eqref{main} for fast work processes. (b) For $T_\mathrm{i}\ll\delta E$, Eq.~\eqref{main1} describes frictional work well irrespective of $\tau$, since the adiabatic state is well-approximated by a two-level system. All results are for $N=8$, $L=g$, $h_\mathrm{i}=1.5g$, and $\Delta h=2g$, with $\delta E\approx 4.98g$.}

    \label{fig:fig3new}
\end{figure}

We assume an initial thermal state $\rho_\mathrm{i}=e^{-\hat{H}_\mathrm{i}/T_\mathrm{i}}/\operatorname{Tr}[e^{-\hat{H}_\mathrm{i}/T_\mathrm{i}}]$, with $\hat{H}_\mathrm{i}$ the initial Hamiltonian with transverse field strength $h_\mathrm{i}$. The system is then driven unitarily over a duration $\tau$ by a time-dependent field $h(t)$, see Fig.~\ref{fig:protocol}. For the results presented here we use a linear ramp $h(t)=h_\mathrm{i}+(t/\tau)\Delta h$, although similar results hold for other drive protocols. The final Hamiltonian is $\hat{H}_\mathrm{f}$ with transverse field strength $h_\mathrm{f}$. Strictly, the minimum work principle assumes no level crossings~\cite{PhysRevE.71.046107}. We take $h_\mathrm{i},h_\mathrm{f} > g$ to avoid effects of the paramagnetic-ferromagnetic crossover at $h\approx g$~\cite{PhysRevB.22.436}. We find that the minimum work principle then holds in the results presented here, despite higher energy-level crossings. We take $\Delta h=h_\mathrm{f}-h_\mathrm{i}>0$, in which case the drive extracts work from the system ($\langle W\rangle_\tau<0$). For $L \neq 0$, the initial state is obtained numerically from exact diagonalisation of $\hat{H}_\mathrm{i}$ and time evolved using Runge-Kutta integration. The final adiabatic state is obtained using a long protocol duration ($\tau=100g^{-1}$).

\begin{figure*}[!]
    \includegraphics[width=\linewidth]{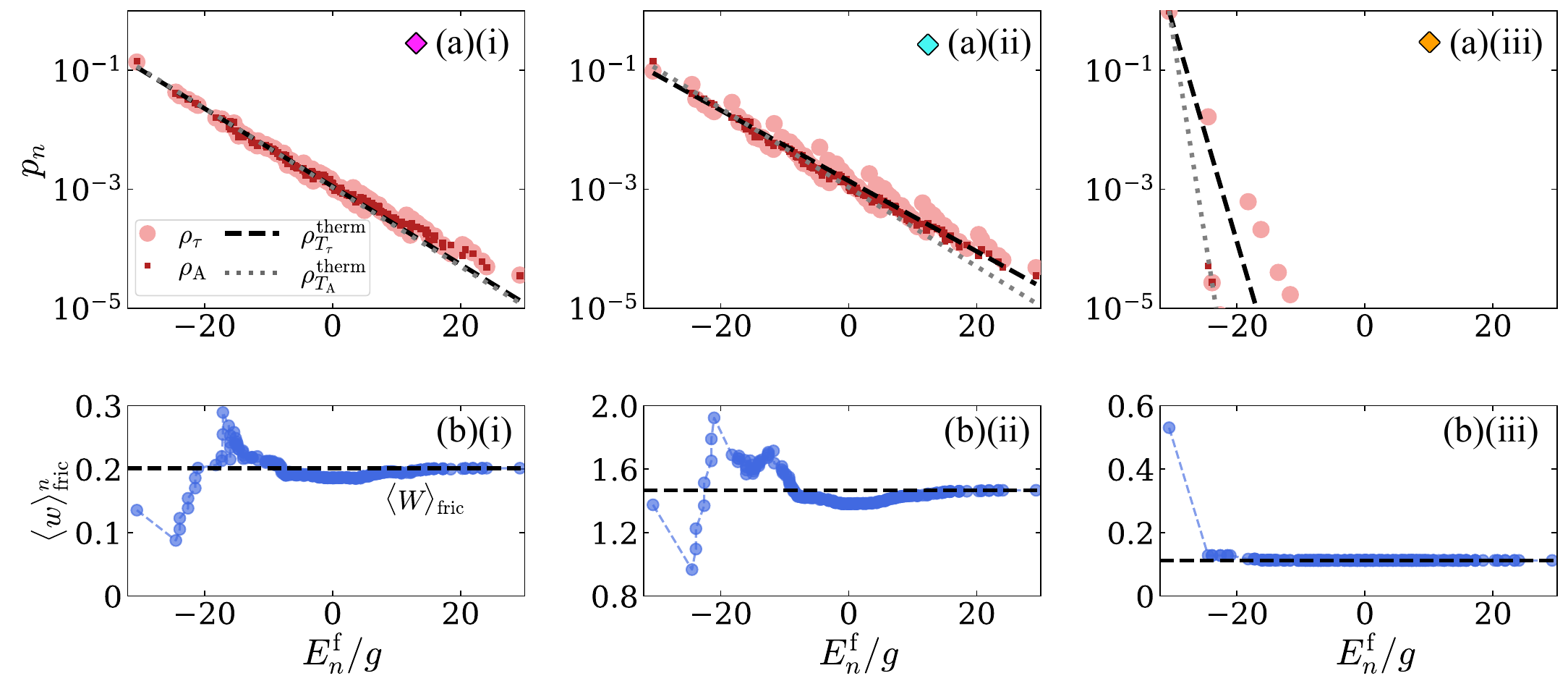} 
\caption{(a) Population distributions $p_n=\langle n_\mathrm{f}|\rho| n_\mathrm{f}\rangle$ in the final energy eigenbasis for $\rho_\tau$ (light red circles), $\rho_{T_\tau}^\mathrm{therm}$ (black dashed line), $\rho_\mathrm{A}$ (dark red squares) and $\rho_{T_\mathrm{A}}^\mathrm{therm}$ (gray dotted line), corresponding to the diamonds in Fig.~\ref{fig:fig2new}. (i) $\tau= g^{-1}$ and $T_\mathrm{i} = 4 g$. Here $T_\tau\approx T_\mathrm{A}$ and $\langle W\rangle_\mathrm{fric}\approx T_\tau\Delta S_\mathrm{d}\approx T_\mathrm{A}\Delta S_\mathrm{d}$ [Eq.~\eqref{main} and Eq.~\eqref{mainTA}]. (ii) $\tau = 0$ and $T_\mathrm{i} = 4 g$. Here $T_\tau\not\approx T_\mathrm{A}$ and $\langle W\rangle_\mathrm{fric}\approx T_\tau\Delta S_\mathrm{d}\not\approx T_\mathrm{A}\Delta S_\mathrm{d}$ [Eq.~\eqref{main}]. (iii) $\tau=1.5g^{-1}$ and $T_\mathrm{i} = 0.5g$. Only the first two levels of $\rho_\mathrm{A}$ are appreciably occupied and $\langle W\rangle_\mathrm{fric}\approx T_\mathrm{A} D(\rho_\tau\|\rho_\mathrm{T_\mathrm{A}}^\mathrm{therm})$ follows from Eq.~\eqref{plastina}. (b) Corresponding cumulative contribution to the frictional work up to a given energy level [see Eq.~\eqref{cumulative}]. All results are for $N=8$, $L=g$, $h_\mathrm{i}=1.5g$ and $\Delta h=2g$.}

\label{fig:fig4pop}
\end{figure*}

\section{Results}

\subsection{Frictional work in a non-integrable spin chain}\label{sec:main}

Our main result Eq.~\eqref{main} is demonstrated in Figure~\ref{fig:fig2new}. Frictional work is well-described by Eq.~\eqref{main} for varying protocol duration $\tau$ [Fig.~\ref{fig:fig2new}(a)] and work step size $\Delta h$ [Fig.~\ref{fig:fig2new}(b)] as long as the temperature of the initial state satisfies $T_\mathrm{i}\gtrsim \delta E$. Here $\delta E$ is the energy gap between the ground and first excited state of $\hat{H}_\mathrm{i}$. For low temperatures $T_\mathrm{i}\ll \delta E$, Eq.~\eqref{main} continues to describe frictional work as long as the work process is slow, $\tau\gtrsim\Delta h/g^2$, see Fig.~\ref{fig:fig2new}(c). Here the parameter $\Delta h/(g^2\tau)$ provides an estimate of the strength of non-adiabatic effects in the evolution~\cite{sinitsyn2016}.\footnote{More precisely, non-adiabatic effects are small when $|\braket{n(t)|\partial_t\hat{H}(t)|m(t)}|/(E_n(t)-E_m(t))^2\ll 1$, with $\ket{n(t)}$ the eigenstates of $\hat{H}(t)$ with corresponding eigenvalues $E_n(t)$. The adiabatic parameter $\Delta h/(g^2\tau)$ is obtained from approximating $|\braket{n(t)|\partial_t\hat{H}(t)|m(t)}|\sim \Delta h/\tau$ and $|E_n(t)-E_m(t)|\sim g$.} The only regime we have identified where Eq.~\eqref{main} does not hold is for fast work processes ($\tau\ll \Delta h/g^2$) starting from a cold initial state ($T_\mathrm{i}<\delta E$). This regime will be discussed shortly.

For slow work processes $\tau\gtrsim \Delta h/g^2$, we find that $T_\tau\approx T_\mathrm{A}$ and Eq.~\eqref{main} reduces to Eq.~\eqref{mainTA}, see Fig.~\ref{fig:fig2new}(a). The validity of Eq.~\eqref{mainTA} can be motivated as follows. In general, the frictional work can be written as
\begin{equation}\label{wfricQE3}
\langle W \rangle_{\mathrm{fric}} =T\Delta S_{\mathrm{d}}+F_T(\rho_\tau^\mathrm{diag})-F_T(\rho_\mathrm{A}).
\end{equation}
Here $F_T(\rho)=\mathrm{Tr}(\rho\hat{H}_\mathrm{f})-T S(\rho)$ is the quantum free energy~\cite{GAVEAU1997347,Skrzypczyk2014,Parrondo2015,PhysRevLett.134.050401} {and $T$ is arbitrary}. The approximation $\langle W\rangle_\mathrm{fric}\approx T\Delta S_\mathrm{d}$ requires $|F_T(\rho_\tau^\mathrm{diag})-F_T(\rho_\mathrm{A})|\ll T\Delta S_{\mathrm d }$. In general, $F_T$ is minimized for a Gibbs state, $\nabla_\rho F_T|_{\rho_T^\mathrm{therm}}=0$, hence $F_T$ varies at second order in deviations from $\rho_T^\mathrm{therm}$.\footnote{Here $\nabla_\rho F_T$ is a matrix with elements $\partial F_T/\partial \rho_{ij}$. To see that $F_T$ is minimized by a Gibbs state, first note that $S(\rho)\le S(\rho^\mathrm{diag})$~\cite{PhysRevLett.113.140401} and hence $F_T(\rho)\ge F_T(\rho^\mathrm{diag})$. Minimising $F_T(\rho^\mathrm{diag})$ with respect to $\rho^\mathrm{diag}$ then gives $\rho^\mathrm{diag}=\rho_T^\mathrm{therm}$.} In comparison, $S_\mathrm{d}$ varies at first order ($\nabla_\rho S_\mathrm{d}|_{\rho_T^\mathrm{therm}}\ne 0$). When both $F(\rho_\mathrm{A})$ and $F(\rho_\tau^\mathrm{diag})$ are in the vicinity of the minimum $F(\rho_T^\mathrm{therm})$, such that $\nabla_\rho F_T|_{\rho_\tau^\mathrm{diag}}\approx\nabla_\rho F_T|_{\rho_\mathrm{A}}\approx 0$, we will have $|F_T(\rho_\tau^\mathrm{diag})-F_T(\rho_\mathrm{A})|\ll T\Delta S_{\mathrm d }$ and $W_\mathrm{fric}\approx T\Delta S_\mathrm{d}$. For slow processes, this occurs for a temperature $T\approx T_\tau\approx T_\mathrm{A}$. Note also that for small $\Delta h$, $T_\tau\approx T_\mathrm{i}$ and Eq.~\eqref{main} follows directly from the result $\langle W\rangle_\mathrm{fric} \approx T_\mathrm{i} \Delta S_\mathrm{d}$ from Ref.~\cite{POLKOVNIKOV2011486}.

Deviations from Eq.~\eqref{main} become appreciable for fast work processes ($\tau\ll\Delta h/g^2$) when the initial temperature is less than the energy gap between the ground and first excited state ($T_\mathrm{i}<\delta E$). Here we find instead that frictional work is well approximated by Eq.~\eqref{main1}, see Fig.~\ref{fig:fig3new}. Using the definition of relative entropy, it is straightforward to show that, for arbitrary $T$,
\begin{equation}\label{wfricQE1}
\langle W \rangle_{\mathrm{fric}} = TD(\rho_\tau \| \rho_T^\mathrm{therm}) - TD(\rho_\mathrm{A} \| \rho_T^\mathrm{therm}).
\end{equation}
The result $\langle W \rangle_{\mathrm{fric}}\approx TD(\rho_\tau \| \rho_T^\mathrm{therm})$ is a good approximation when $D(\rho_\tau \| \rho_T^\mathrm{therm})\gg D(\rho_\mathrm{A} \| \rho_T^\mathrm{therm})$, which in terms of quantum free energy is $F_T(\rho_\tau)-F_T(\rho_T^\mathrm{therm})\gg F_T(\rho_\mathrm{A})-F_T(\rho_T^\mathrm{therm})$. {The term $TD(\rho_\mathrm{A} \| \rho_T^\mathrm{therm})$ is minimized for a temperature $T'$ that satisfies $\operatorname{Tr}[\rho_{T'}^\mathrm{therm}\ln\rho_{T'}^\mathrm{therm}]=\operatorname{Tr}[\rho_\mathrm{A}\ln\rho_\mathrm{A}]$, which we find is very close to $T_\mathrm{A}$ defined by energy matching ($|T'-T_\mathrm{A}|/T'\lesssim 10^{-2}$)}. Typically, $D(\rho_\tau \| \rho_{T_\mathrm{A}}^\mathrm{therm})\gg D(\rho_\mathrm{A} \| \rho_{T_\mathrm{A}}^\mathrm{therm})$ for fast work processes at low temperatures, as such processes generate substantial coherences that give $T_\tau\gg T_\mathrm{A}$. At very low temperatures $(T_\mathrm{i}\ll\delta E)$, occupation beyond the first two energy levels of $\rho_\mathrm{A}$ are frozen out and $\rho_\mathrm{A}$ can be well-approximated by a two-level Gibbs state~\cite{PhysRevB.109.024310}, in which case Eq.~\eqref{plastina} can be applied directly {irrespective of $\tau$ [see Fig.~\ref{fig:fig3new}(b)]}.

Figure~\ref{fig:fig4pop}(a) plots energy-level distributions $p_n=\langle n_\mathrm{f}|\rho|n_\mathrm{f}\rangle$ for $\rho_\tau$ and $\rho_\mathrm{A}$ for the three parameter regimes indicated by diamonds in Fig.~\ref{fig:fig2new}. Here $\ket{n_\mathrm{f}}$ are the final energy eigenstates with corresponding eigenvalues $E_n^\mathrm{f}$. For $\tau\gtrsim \Delta h/g^2$, the distributions for $\rho_\mathrm{A}$ and $\rho_\tau$ are similar [Fig.~\ref{fig:fig4pop}(a)(i)] and frictional work is described by Eq.~\eqref{mainTA}. Deviation between $\rho_\mathrm{A}$ and $\rho_\tau$ become appreciable for $\tau<\Delta h/g^2$ [Fig.~\ref{fig:fig4pop}(a)(ii)], in which case $T_\tau\not\approx T_\mathrm{A}$  and frictional work is described by Eq.~\eqref{main}. At very low temperatures [Fig.~\ref{fig:fig4pop}(iii)], $\rho_\mathrm{A}$ can be well-approximated by a two-level Gibbs state and Eq.~\eqref{plastina} can be applied directly.

Figure~\ref{fig:fig4pop}(a) also compares the energy-level distributions $p_n$ for $\rho_\tau$ and $\rho_\mathrm{A}$ with their corresponding Gibbs-state distributions. For reference, the contribution to the frictional work up to a given energy level $n$, 
\begin{equation}\label{cumulative}
    \langle w\rangle_\mathrm{fric}^n\equiv\sum_{k=1}^n [p_k(\rho_\tau)-p_k(\rho_\mathrm{A})]E_k^\mathrm{f},
\end{equation}
is shown in Fig.~\ref{fig:fig4pop}(b). The Gibbs state $\rho_{T_\mathrm{A}}^\mathrm{therm}$ closely follows $\rho_\mathrm{A}$ for energy levels where $|\langle w\rangle_\mathrm{fric}^n|$ is appreciable. Although $\rho_\tau$ deviates from a Gibbs state, its average behavior is still well-described by $\rho_{T_\tau}^\mathrm{therm}$.

\subsection{Comparison with frictional work in an integrable spin chain}\label{sec:integrable}

In the absence of a longitudinal field, $L=0$, Eq.~\eqref{eq:H} is integrable can be diagonlized by mapping onto $N$ non-interacting fermions~\cite{PFEUTY197079,mbeng2020quantum,Sachdev_2011},
\begin{equation}
    \hat{H}(t)=E_0+\sum_{j=1}^N\omega_j\hat{c}_j^\dagger\hat{c}_j.
    \label{fermion}
\end{equation}
Here $\hat{c}_j$ is the lowering operator for the $j$th fermion,
\begin{equation}\label{spectra}
    \omega_j\equiv 2\sqrt{h^2+g^2-2gh\cos\theta_j},\hspace{0.5cm}\theta_j\equiv \frac{(2j-1)\pi}{N},
\end{equation} 
 and $E_0\equiv-\frac{1}{2}\sum_{j=1}^N\omega_j$ is the ground-state energy. Each fermion in Eq.~\eqref{fermion} evolves as an independent two-level system $\rho_\tau^j$~\cite{dziarmaga2005,mbeng2020quantum}, resulting in a separable time-evolved density matrix $\rho_\tau=\rho_\tau^1\otimes\rho_\tau^2\otimes...\otimes\rho_\tau^N$. The total frictional work is then $\langle W\rangle_\mathrm{fric}=\sum_{j=1}^N\langle W\rangle_\mathrm{fric}^j$, with $\langle W \rangle_\mathrm{fric}^j$ the contribution from fermion $j$.
 
\begin{figure*}[!]
    \centering
    \includegraphics[width=\linewidth]{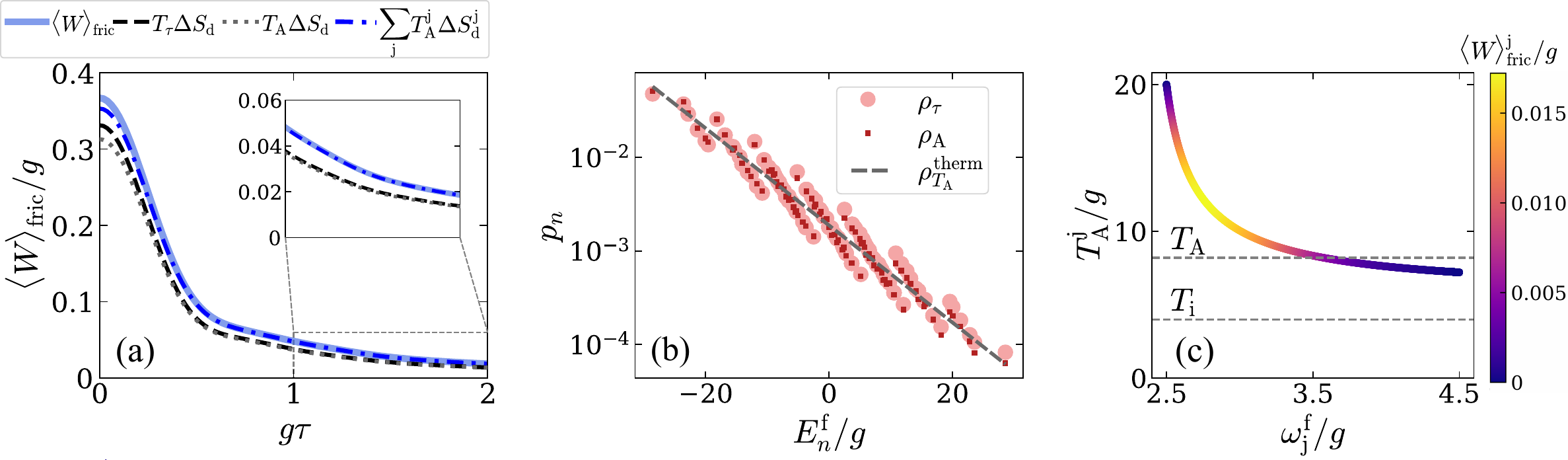}
    \caption{(a) Frictional work (light blue solid line) for an integrable spin chain ($N=8$) is poorly described by Eq.~\eqref{main} (dashed black line) but is well described by the modified expression Eq.~\eqref{eq:WDeltaS} (dark blue dot-dashed line). (b) Corresponding population distributions $p_n$ in the final energy eigenbasis for $\rho_\tau$ (light red circles), $\rho_\mathrm{A}$ (dark red squares) and $\rho_{T_\mathrm{A}}^\mathrm{therm}$ (gray dashed line) for $\tau=g^{-1}$. (c) The effective temperature $T_\mathrm{A}^j$ for each fermion in a large ($N=5000$) integrable spin chain with $\tau=g^{-1}$. The temperature of each mode is plotted against the final fermion energy $\omega_j^\mathrm{f}$, with the color indicating the contribution of frictional work from each mode. All results are for $L=0$, $T_\mathrm{i}=4g$, $h_\mathrm{i}=1.5g$ and $\Delta h =2g$.}
    \label{fig:fig4}
\end{figure*}

\begin{figure}[!]
    \centering
    \includegraphics[width=\linewidth]{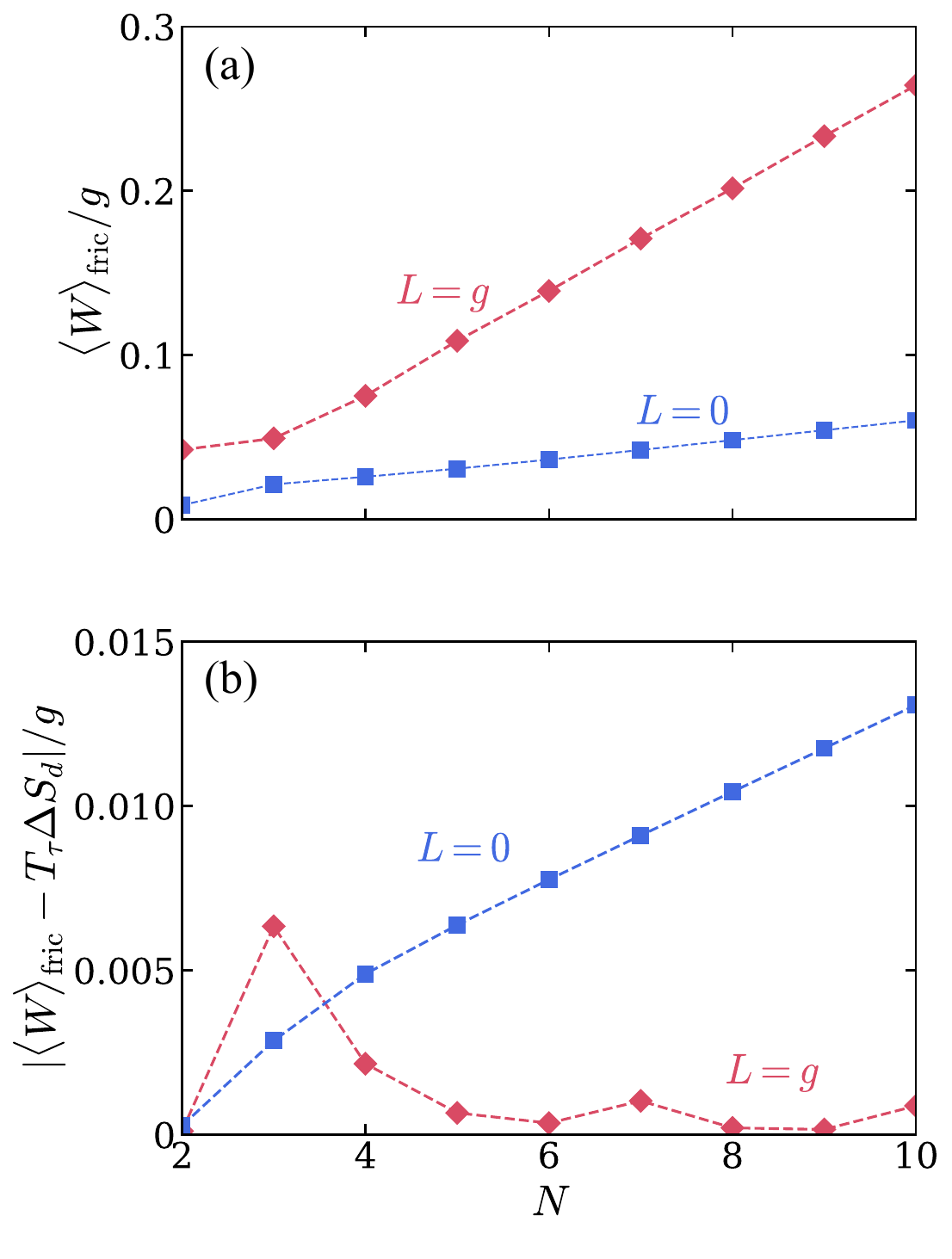}
    \caption{(a) Frictional work grows extensively as the number of spins $N$ in the chain increases for both $L=0$ (blue squares) and $L=g$ (red diamonds). (b) Deviation from Eq.~\eqref{main}, $|\langle W\rangle_\mathrm{fric}-T_\tau\Delta S_\mathrm{d}|$, grows extensively for $L=0$, whereas it ceases to grow beyond $N=3$ for $L=g$. All results are for $T_\mathrm{i}=4g$, $h_\mathrm{i}=1.5g$, $\Delta h =2g$ and $\tau=g^{-1}$.}
    \label{fig:fig7}
\end{figure}

\begin{figure}[!]
\centering
\includegraphics[width=\linewidth]{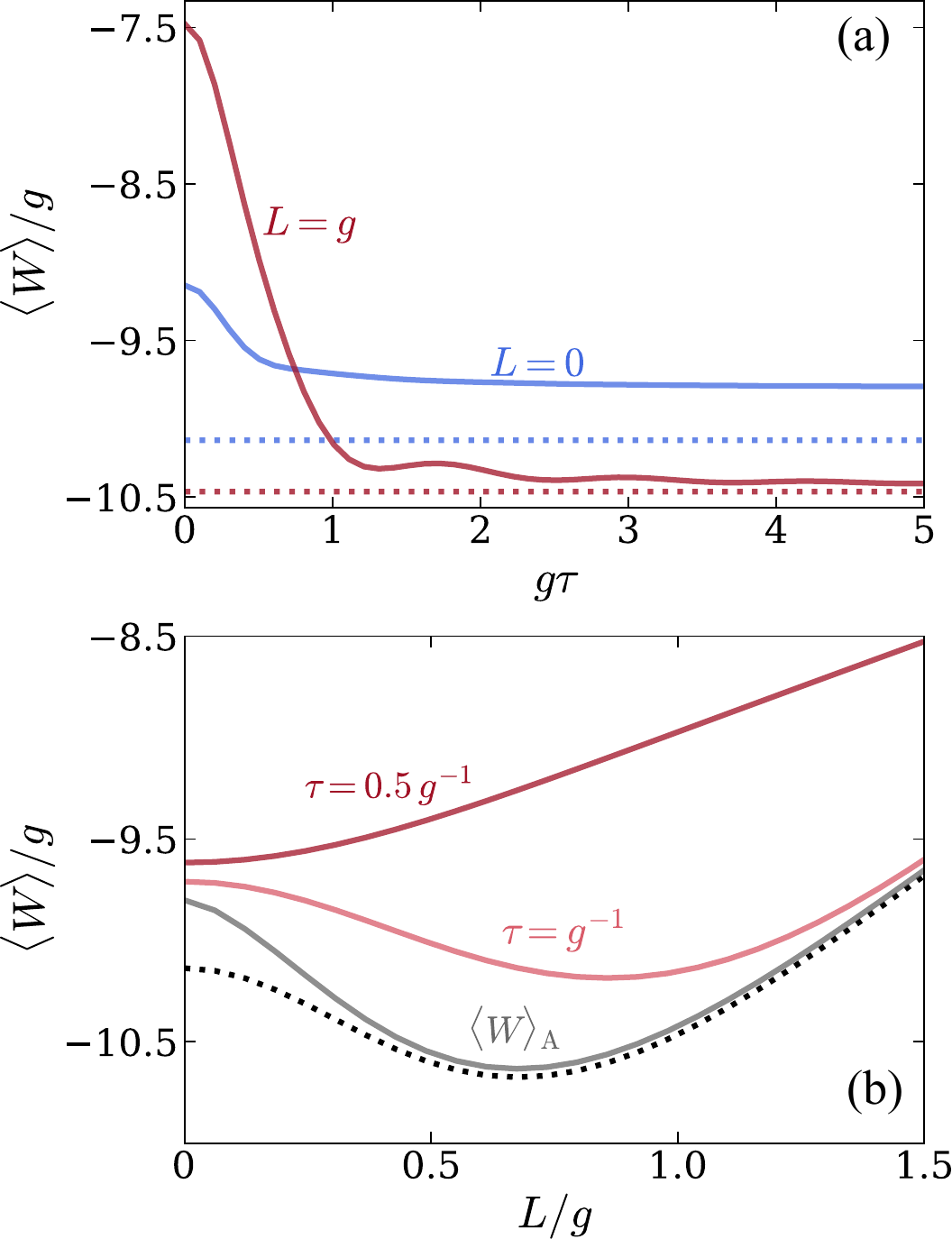}
\caption{(a) Work (solid lines) extracted from an integrable ($L=0$, blue line) and non-integrable ($L=g$, red line) as a function of extraction duration $\tau$. Matching colored dotted lines are the corresponding optimal work outputs $\langle W \rangle_\mathrm{opt}$ as defined by Eq.~\eqref{Wopt}. For adiabatic extraction, integrability breaking reduces $\langle W\rangle_\tau-\langle W\rangle_\mathrm{opt}$, since $\rho_\mathrm{A}$ is well approximated by $\rho_{T_\mathrm{A}}^\mathrm{therm}$. For fast processes ($\tau\ll \Delta h/g^2$), integrability breaking gives rise to larger frictional work and hence increases $\langle W\rangle_\tau-\langle W\rangle_\mathrm{opt}$ compared to the integrable case. (b) Work (solid lines) as a function of the longitudinal field $L$ for different $\tau$ compared with $\langle W\rangle_\mathrm{opt}$ (black dotted line). All results are for $N=8$, $T_\mathrm{i} =2g$, $h_\mathrm{i}= 1.5g$ and $\Delta h = 2g$. Qualitatively similar results for $\langle W\rangle_\tau-\langle W\rangle_\mathrm{opt}$ are obtained at larger $T_i$ also.}

\label{fig:finalfigure}
\end{figure}

Compared to the non-integrable spin chain, the frictional work is no longer well approximated by Eq.~\eqref{main} or Eq.~\eqref{mainTA}, see Fig.~\ref{fig:fig4}(a). Since each fermion evolves as an independent two-level system, and any two-level system diagonal in the energy eigenbasis is thermal~\cite{quan2007}, we can use Eq.~\eqref{plastina} to describe each $\langle W \rangle_\mathrm{fric}^j$. This gives
\begin{equation}\label{fermionfric}
    \langle W \rangle_\mathrm{fric}=\sum_{j=1}^NT^j_\mathrm{A}D(\rho_\tau^j\|\rho_\mathrm{A}^j).
\end{equation}
Here $\rho_\mathrm{A}^j$ is the adiabatic limit of $\rho_\tau^j$ and is a Gibbs state with temperature $T_\mathrm{A}^j$. Using Eq.~\eqref{eq:Ddecom}, the relative entropy can be expanded for large $\tau$ to give 
\begin{equation}\label{eq:WDeltaS}
    \langle W \rangle_\mathrm{fric}=\sum_{j=1}^NT^j_\mathrm{A}\Delta S^j_\mathrm{d}+O\left(\frac{1}{\tau^2}\right).
\end{equation}
with $\Delta S_\mathrm{d}^j$ the contribution to the diagonal entropy from fermion $j$. The result Eq.~\eqref{eq:WDeltaS} describes the frictional work well over a broad range of $\tau$, only deviating for very small $\tau$, see Fig.~\ref{fig:fig4}(a).

The difference between Eq.~\eqref{eq:WDeltaS} and Eq.~\eqref{mainTA} arises from the variation in $T^j_\mathrm{A}$, such that $\rho_\mathrm{A}$ {deviates substantially from a Gibbs state}, see Fig.~\ref{fig:fig4}(b). The individual $T_\mathrm{A}^j$ for a large-$N$ system are shown in Fig.~\ref{fig:fig4}(c) as a function of the final fermionic energies $\omega_j^\mathrm{f}$, along with the contribution to the frictional work from each fermion. Notably, the temperature range of fermions that contribute appreciably to the frictional work is large ($6g\lesssim T_\mathrm{A}^j\lesssim 14g$) compared to $T_\mathrm{A}\approx 6g$. {This is not surprising, as the absence of interactions between the fermions prevents their mutual thermalization to a common temperature. More generally, we expect a similar absence of a common temperature even in integrable systems that cannot be mapped onto a non-interacting system, as the extensive number of conserved quantities constrains the dynamics.}\footnote{{Analogously, an isolated integrable system fails to equilibrate to a Gibbs ensemble and instead equilibrates to a generalized Gibbs ensemble~\cite{rigol2007}.}} The addition of a longitudinal field $L$ results in interactions between the fermions and a single effective temperature $T_\mathrm{A}$ describes $\rho_\mathrm{A}$, as demonstrated in Fig.~\ref{fig:fig4pop}(a).

The significance of integrability breaking is further demonstrated in Fig.~\ref{fig:fig7}, which shows frictional work and deviations from Eq.~\eqref{main} as a function of spin number $N$. Irrespective of $L$, frictional work grows extensively [Fig.~\ref{fig:fig7}(a)]. Deviation from Eq.~\eqref{main} grows extensively for $L=0$ but ceases to grow beyond $N\approx 3$ for $L=g$ [Fig.~\ref{fig:fig7}(b)]. As a result, the relative error $|\langle W\rangle_\mathrm{fric}-T_\tau\Delta S_\mathrm{d}|/\langle W\rangle_\mathrm{fric}$ approaches a non-zero constant for $L=0$, whereas it decreases with increasing $N$ for $L=g$.

\subsection{Impact of integrability on total work extraction}\label{non_integrable}
In this final subsection we characterize the effect of integrability on total work output. To enable comparison between the integrable and non-integrable results, we introduce the optimal work output
\begin{equation}\label{Wopt}
    \langle W\rangle_\mathrm{opt}\equiv\operatorname{Tr}[\rho_{T_\mathrm{A}}^\mathrm{therm}\hat{H}_\mathrm{f}]-\operatorname{Tr}[\rho_\mathrm{i}\hat{H}_\mathrm{i}].
\end{equation}
The state $\rho_{T_\mathrm{A}}^\mathrm{therm}$ minimizes the energy over all states with fixed diagonal entropy~\cite{Allahverdyan2004} and therefore $\langle W\rangle_\tau\ge\langle W\rangle_\mathrm{A}\ge\langle W\rangle_\mathrm{opt}$ (with $\langle W\rangle_\mathrm{A}=\langle W\rangle_\mathrm{opt}$ when $\rho_\mathrm{A}=\rho_{T_\mathrm{A}}^\mathrm{therm}$). {The difference $\langle W\rangle_\tau-\langle W\rangle_\mathrm{opt}$ is the maximum work (per copy) that can be extracted by a cyclic unitary acting on multiple copies $\lim_{K\rightarrow\infty}\rho_\tau^{\otimes K}$ of $\rho_\tau$~\cite{pusz1978,alicki2013}. This difference therefore gives an indication of how close $\langle W\rangle_\mathrm{\tau}$ is to optimal.}

Figure~\ref{fig:finalfigure}(a) compares $\langle W\rangle_\tau$ to $\langle W\rangle_\mathrm{opt}$ for both the integrable ($L=0$) and the non-integrable ($L=g$) spin chains for different work durations $\tau$. In the adiabatic limit, $\langle W\rangle_\tau-\langle W\rangle_\mathrm{opt}$ is smaller in the non-integrable system, as $\rho_\mathrm{A}$ is better approximated by a thermal state. In contrast, for rapid work extraction ($\tau \ll \Delta h/g^2$), both $\langle W\rangle_\tau-\langle W\rangle_\mathrm{opt}$ and $\langle W\rangle_\mathrm{fric}$ are smaller in the integrable system. The suppression of frictional work in the integrable system is due to conservation laws restricting non-adiabatic transitions (this behavior is analogous to classical systems, where integrability confines motion to regions within the total phase space, reducing the accessible volume for the dynamics~\cite{D'Alessio03052016}). This suppression is also clear in Fig.~\ref{fig:fig4}(b), as $p_n(\rho_\tau)$ remains close to $p_n(\rho_\mathrm{A})$ even for rapid work extraction. Figure~\ref{fig:finalfigure}(b) compares $\langle W\rangle_\tau$ and $\langle W\rangle_\mathrm{opt}$ as a function of $L$, and shows that $\langle W\rangle_\tau-\langle W\rangle_\mathrm{opt}$ decreases with $L$ for $\tau\gtrsim \Delta h/g^2$, whereas it increases with $L$ for $\tau\ll \Delta h/g^2$.

\section{Conclusion}\label{conclusion}
 
We have extended the results of Plastina \emph{et al}.\ by showing that frictional work in a non-integrable spin chain is well-described by diagonal entropy production, despite the adiabatically-evolved state deviating from a Gibbs state. The relation is characterized by the effective temperature of the final time-evolved state and holds for a range of initial temperatures and work-processes. The relation breaks down for low temperatures and fast work processes, in which case frictional work is instead described by the relative entropy between the time-evolved state and a thermal state at the effective temperature of the adiabatic state. We compare our results to those in an integrable spin chain, in which case a single effective temperature no longer describes the system and frictional work is instead characterized by a sum of contributions from each independent subspace of the system. We show that integrability breaking enhances work output in the adiabatic limit but degrades work output for sufficiently rapid work extraction.

It would be interesting to extend the analysis presented here to other non-integrable systems, particularly those with integrable limits that cannot be mapped onto a non-interacting system. The one-dimensional Bose gas is one such example, which is integrable in the homogeneous Lieb-Liniger regime~\cite{cazalilla2011}. Work extraction from this system has recently been studied using various engine cycles~\cite{WatsonKheruntsyan2025,watson2025,nautiyal2024,PhysRevE.111.054133}. {Although we have confined our analysis to work processes starting from an initial thermal state, it would also be interesting to consider initial passive states that are non-thermal, such that both the initial and final states of the work process are described by effective temperatures.} An extension of the results presented here could explore effects of integrability breaking on frictional work fluctuations~\cite{PhysRevLett.113.260601}, which are more sensitive to non-equilibrium aspects of the underlying ensemble~\cite{jeon2015,PhysRevE.94.012125,PhysRevE.111.L012102,gong2015,hoang2018,alhambra2016}. Finally, the impact of integrability breaking on the performance of quantum thermal machine cycles could be explored. Here performance depends not only on the work output but also the heat exchanged with the reservoir, which will likely be sensitive to the presence or absence of integrability.

\begin{acknowledgments}

This research was supported by The University of Queensland--IITD Academy of Research (UQIDAR), the Australian Research Council Centre of Excellence for Engineered Quantum Systems (CE170100009), the Australian Research Council Discovery Project DP260103158, and the Australian federal government Department of Industry, Science, and Resources via the Australia-India Strategic Research Fund (AIRXIV000025). We thank Andrew Groszek for comments on the manuscript.
\end{acknowledgments}


\begin{thebibliography}{74}%
\makeatletter
\providecommand \@ifxundefined [1]{%
 \@ifx{#1\undefined}
}%
\providecommand \@ifnum [1]{%
 \ifnum #1\expandafter \@firstoftwo
 \else \expandafter \@secondoftwo
 \fi
}%
\providecommand \@ifx [1]{%
 \ifx #1\expandafter \@firstoftwo
 \else \expandafter \@secondoftwo
 \fi
}%
\providecommand \natexlab [1]{#1}%
\providecommand \enquote  [1]{``#1''}%
\providecommand \bibnamefont  [1]{#1}%
\providecommand \bibfnamefont [1]{#1}%
\providecommand \citenamefont [1]{#1}%
\providecommand \href@noop [0]{\@secondoftwo}%
\providecommand \href [0]{\begingroup \@sanitize@url \@href}%
\providecommand \@href[1]{\@@startlink{#1}\@@href}%
\providecommand \@@href[1]{\endgroup#1\@@endlink}%
\providecommand \@sanitize@url [0]{\catcode `\\12\catcode `\$12\catcode `\&12\catcode `\#12\catcode `\^12\catcode `\_12\catcode `\%12\relax}%
\providecommand \@@startlink[1]{}%
\providecommand \@@endlink[0]{}%
\providecommand \url  [0]{\begingroup\@sanitize@url \@url }%
\providecommand \@url [1]{\endgroup\@href {#1}{\urlprefix }}%
\providecommand \urlprefix  [0]{URL }%
\providecommand \Eprint [0]{\href }%
\providecommand \doibase [0]{https://doi.org/}%
\providecommand \selectlanguage [0]{\@gobble}%
\providecommand \bibinfo  [0]{\@secondoftwo}%
\providecommand \bibfield  [0]{\@secondoftwo}%
\providecommand \translation [1]{[#1]}%
\providecommand \BibitemOpen [0]{}%
\providecommand \bibitemStop [0]{}%
\providecommand \bibitemNoStop [0]{.\EOS\space}%
\providecommand \EOS [0]{\spacefactor3000\relax}%
\providecommand \BibitemShut  [1]{\csname bibitem#1\endcsname}%
\let\auto@bib@innerbib\@empty
\bibitem [{\citenamefont {Vinjanampathy}\ and\ \citenamefont {Anders}(2016)}]{vinjanampathy2016}%
  \BibitemOpen
  \bibfield  {author} {\bibinfo {author} {\bibfnamefont {S.}~\bibnamefont {Vinjanampathy}}\ and\ \bibinfo {author} {\bibfnamefont {J.}~\bibnamefont {Anders}},\ }\bibfield  {title} {\bibinfo {title} {Quantum thermodynamics},\ }\href {https://doi.org/10.1080/00107514.2016.1201896} {\bibfield  {journal} {\bibinfo  {journal} {Contemp. Phys.}\ }\textbf {\bibinfo {volume} {57}},\ \bibinfo {pages} {545} (\bibinfo {year} {2016})}\BibitemShut {NoStop}%
\bibitem [{\citenamefont {Uzdin}\ \emph {et~al.}(2015)\citenamefont {Uzdin}, \citenamefont {Levy},\ and\ \citenamefont {Kosloff}}]{uzdin2015}%
  \BibitemOpen
  \bibfield  {author} {\bibinfo {author} {\bibfnamefont {R.}~\bibnamefont {Uzdin}}, \bibinfo {author} {\bibfnamefont {A.}~\bibnamefont {Levy}},\ and\ \bibinfo {author} {\bibfnamefont {R.}~\bibnamefont {Kosloff}},\ }\bibfield  {title} {\bibinfo {title} {Equivalence of quantum heat machines, and quantum-thermodynamic signatures},\ }\href {https://doi.org/10.1103/PhysRevX.5.031044} {\bibfield  {journal} {\bibinfo  {journal} {Phys. Rev. X}\ }\textbf {\bibinfo {volume} {5}},\ \bibinfo {pages} {031044} (\bibinfo {year} {2015})}\BibitemShut {NoStop}%
\bibitem [{\citenamefont {Klatzow}\ \emph {et~al.}(2019)\citenamefont {Klatzow}, \citenamefont {Becker}, \citenamefont {Ledingham}, \citenamefont {Weinzetl}, \citenamefont {Kaczmarek}, \citenamefont {Saunders}, \citenamefont {Nunn}, \citenamefont {Walmsley}, \citenamefont {Uzdin},\ and\ \citenamefont {Poem}}]{klatzow2019}%
  \BibitemOpen
  \bibfield  {author} {\bibinfo {author} {\bibfnamefont {J.}~\bibnamefont {Klatzow}}, \bibinfo {author} {\bibfnamefont {J.~N.}\ \bibnamefont {Becker}}, \bibinfo {author} {\bibfnamefont {P.~M.}\ \bibnamefont {Ledingham}}, \bibinfo {author} {\bibfnamefont {C.}~\bibnamefont {Weinzetl}}, \bibinfo {author} {\bibfnamefont {K.~T.}\ \bibnamefont {Kaczmarek}}, \bibinfo {author} {\bibfnamefont {D.~J.}\ \bibnamefont {Saunders}}, \bibinfo {author} {\bibfnamefont {J.}~\bibnamefont {Nunn}}, \bibinfo {author} {\bibfnamefont {I.~A.}\ \bibnamefont {Walmsley}}, \bibinfo {author} {\bibfnamefont {R.}~\bibnamefont {Uzdin}},\ and\ \bibinfo {author} {\bibfnamefont {E.}~\bibnamefont {Poem}},\ }\bibfield  {title} {\bibinfo {title} {Experimental demonstration of quantum effects in the operation of microscopic heat engines},\ }\href {https://doi.org/10.1103/PhysRevLett.122.110601} {\bibfield  {journal} {\bibinfo  {journal} {Phys. Rev. Lett.}\ }\textbf {\bibinfo {volume} {122}},\ \bibinfo {pages} {110601} (\bibinfo {year}
  {2019})}\BibitemShut {NoStop}%
\bibitem [{\citenamefont {Korzekwa}\ \emph {et~al.}(2016)\citenamefont {Korzekwa}, \citenamefont {Lostaglio}, \citenamefont {Oppenheim},\ and\ \citenamefont {Jennings}}]{korzekwa2016}%
  \BibitemOpen
  \bibfield  {author} {\bibinfo {author} {\bibfnamefont {K.}~\bibnamefont {Korzekwa}}, \bibinfo {author} {\bibfnamefont {M.}~\bibnamefont {Lostaglio}}, \bibinfo {author} {\bibfnamefont {J.}~\bibnamefont {Oppenheim}},\ and\ \bibinfo {author} {\bibfnamefont {D.}~\bibnamefont {Jennings}},\ }\bibfield  {title} {\bibinfo {title} {The extraction of work from quantum coherence},\ }\href {https://doi.org/10.1088/1367-2630/18/2/023045} {\bibfield  {journal} {\bibinfo  {journal} {New J. Phys.}\ }\textbf {\bibinfo {volume} {18}},\ \bibinfo {pages} {023045} (\bibinfo {year} {2016})}\BibitemShut {NoStop}%
\bibitem [{\citenamefont {Kammerlander}\ and\ \citenamefont {Anders}(2016)}]{kammerlander2016}%
  \BibitemOpen
  \bibfield  {author} {\bibinfo {author} {\bibfnamefont {P.}~\bibnamefont {Kammerlander}}\ and\ \bibinfo {author} {\bibfnamefont {J.}~\bibnamefont {Anders}},\ }\bibfield  {title} {\bibinfo {title} {Coherence and measurement in quantum thermodynamics},\ }\href {https://doi.org/10.1038/srep22174} {\bibfield  {journal} {\bibinfo  {journal} {Sci. Rep.}\ }\textbf {\bibinfo {volume} {6}},\ \bibinfo {pages} {22174} (\bibinfo {year} {2016})}\BibitemShut {NoStop}%
\bibitem [{\citenamefont {Francica}\ \emph {et~al.}(2020)\citenamefont {Francica}, \citenamefont {Binder}, \citenamefont {Guarnieri}, \citenamefont {Mitchison}, \citenamefont {Goold},\ and\ \citenamefont {Plastina}}]{francica2020}%
  \BibitemOpen
  \bibfield  {author} {\bibinfo {author} {\bibfnamefont {G.}~\bibnamefont {Francica}}, \bibinfo {author} {\bibfnamefont {F.~C.}\ \bibnamefont {Binder}}, \bibinfo {author} {\bibfnamefont {G.}~\bibnamefont {Guarnieri}}, \bibinfo {author} {\bibfnamefont {M.~T.}\ \bibnamefont {Mitchison}}, \bibinfo {author} {\bibfnamefont {J.}~\bibnamefont {Goold}},\ and\ \bibinfo {author} {\bibfnamefont {F.}~\bibnamefont {Plastina}},\ }\bibfield  {title} {\bibinfo {title} {Quantum coherence and ergotropy},\ }\href {https://doi.org/10.1103/PhysRevLett.125.180603} {\bibfield  {journal} {\bibinfo  {journal} {Phys. Rev. Lett.}\ }\textbf {\bibinfo {volume} {125}},\ \bibinfo {pages} {180603} (\bibinfo {year} {2020})}\BibitemShut {NoStop}%
\bibitem [{\citenamefont {Williamson}\ \emph {et~al.}(2025)\citenamefont {Williamson}, \citenamefont {Cerisola}, \citenamefont {Anders},\ and\ \citenamefont {Davis}}]{williamson2024}%
  \BibitemOpen
  \bibfield  {author} {\bibinfo {author} {\bibfnamefont {L.~A.}\ \bibnamefont {Williamson}}, \bibinfo {author} {\bibfnamefont {F.}~\bibnamefont {Cerisola}}, \bibinfo {author} {\bibfnamefont {J.}~\bibnamefont {Anders}},\ and\ \bibinfo {author} {\bibfnamefont {M.~J.}\ \bibnamefont {Davis}},\ }\bibfield  {title} {\bibinfo {title} {Extracting work from coherence in a two-mode bose–einstein condensate},\ }\href {https://doi.org/10.1088/2058-9565/ad8fc9} {\bibfield  {journal} {\bibinfo  {journal} {Quantum Sci. Technol.}\ }\textbf {\bibinfo {volume} {10}},\ \bibinfo {pages} {015040} (\bibinfo {year} {2025})}\BibitemShut {NoStop}%
\bibitem [{\citenamefont {Uzdin}(2016)}]{uzdin2016}%
  \BibitemOpen
  \bibfield  {author} {\bibinfo {author} {\bibfnamefont {R.}~\bibnamefont {Uzdin}},\ }\bibfield  {title} {\bibinfo {title} {Coherence-induced reversibility and collective operation of quantum heat machines via coherence recycling},\ }\href {https://doi.org/10.1103/PhysRevApplied.6.024004} {\bibfield  {journal} {\bibinfo  {journal} {Phys. Rev. Appl.}\ }\textbf {\bibinfo {volume} {6}},\ \bibinfo {pages} {024004} (\bibinfo {year} {2016})}\BibitemShut {NoStop}%
\bibitem [{\citenamefont {Scully}\ \emph {et~al.}(2003)\citenamefont {Scully}, \citenamefont {Zubairy}, \citenamefont {Agarwal},\ and\ \citenamefont {Walther}}]{scully2003}%
  \BibitemOpen
  \bibfield  {author} {\bibinfo {author} {\bibfnamefont {M.~O.}\ \bibnamefont {Scully}}, \bibinfo {author} {\bibfnamefont {M.~S.}\ \bibnamefont {Zubairy}}, \bibinfo {author} {\bibfnamefont {G.~S.}\ \bibnamefont {Agarwal}},\ and\ \bibinfo {author} {\bibfnamefont {H.}~\bibnamefont {Walther}},\ }\bibfield  {title} {\bibinfo {title} {Extracting work from a single heat bath via vanishing quantum coherence},\ }\href {https://doi.org/10.1126/science.1078955} {\bibfield  {journal} {\bibinfo  {journal} {Science}\ }\textbf {\bibinfo {volume} {299}},\ \bibinfo {pages} {862} (\bibinfo {year} {2003})}\BibitemShut {NoStop}%
\bibitem [{\citenamefont {Shastri}\ and\ \citenamefont {Venkatesh}(2024)}]{PhysRevE.109.014102}%
  \BibitemOpen
  \bibfield  {author} {\bibinfo {author} {\bibfnamefont {R.}~\bibnamefont {Shastri}}\ and\ \bibinfo {author} {\bibfnamefont {B.~P.}\ \bibnamefont {Venkatesh}},\ }\bibfield  {title} {\bibinfo {title} {Controlling work output and coherence in finite-time quantum otto engines through monitoring},\ }\href {https://doi.org/10.1103/PhysRevE.109.014102} {\bibfield  {journal} {\bibinfo  {journal} {Phys. Rev. E}\ }\textbf {\bibinfo {volume} {109}},\ \bibinfo {pages} {014102} (\bibinfo {year} {2024})}\BibitemShut {NoStop}%
\bibitem [{\citenamefont {Horodecki}\ and\ \citenamefont {Oppenheim}(2013)}]{horodecki2013}%
  \BibitemOpen
  \bibfield  {author} {\bibinfo {author} {\bibfnamefont {M.}~\bibnamefont {Horodecki}}\ and\ \bibinfo {author} {\bibfnamefont {J.}~\bibnamefont {Oppenheim}},\ }\bibfield  {title} {\bibinfo {title} {Fundamental limitations for quantum and nanoscale thermodynamics},\ }\href {https://doi.org/10.1038/ncomms3059} {\bibfield  {journal} {\bibinfo  {journal} {Nat. Commun.}\ }\textbf {\bibinfo {volume} {4}},\ \bibinfo {pages} {2059} (\bibinfo {year} {2013})}\BibitemShut {NoStop}%
\bibitem [{\citenamefont {Kosloff}\ and\ \citenamefont {Feldmann}(2002)}]{kosloff2002}%
  \BibitemOpen
  \bibfield  {author} {\bibinfo {author} {\bibfnamefont {R.}~\bibnamefont {Kosloff}}\ and\ \bibinfo {author} {\bibfnamefont {T.}~\bibnamefont {Feldmann}},\ }\bibfield  {title} {\bibinfo {title} {Discrete four-stroke quantum heat engine exploring the origin of friction},\ }\href {https://doi.org/10.1103/PhysRevE.65.055102} {\bibfield  {journal} {\bibinfo  {journal} {Phys. Rev. E}\ }\textbf {\bibinfo {volume} {65}},\ \bibinfo {pages} {055102} (\bibinfo {year} {2002})}\BibitemShut {NoStop}%
\bibitem [{\citenamefont {Feldmann}\ and\ \citenamefont {Kosloff}(2003)}]{feldmann2003}%
  \BibitemOpen
  \bibfield  {author} {\bibinfo {author} {\bibfnamefont {T.}~\bibnamefont {Feldmann}}\ and\ \bibinfo {author} {\bibfnamefont {R.}~\bibnamefont {Kosloff}},\ }\bibfield  {title} {\bibinfo {title} {Quantum four-stroke heat engine: Thermodynamic observables in a model with intrinsic friction},\ }\href {https://doi.org/10.1103/PhysRevE.68.016101} {\bibfield  {journal} {\bibinfo  {journal} {Phys. Rev. E}\ }\textbf {\bibinfo {volume} {68}},\ \bibinfo {pages} {016101} (\bibinfo {year} {2003})}\BibitemShut {NoStop}%
\bibitem [{\citenamefont {Feldmann}\ and\ \citenamefont {Kosloff}(2006)}]{feldmann2006}%
  \BibitemOpen
  \bibfield  {author} {\bibinfo {author} {\bibfnamefont {T.}~\bibnamefont {Feldmann}}\ and\ \bibinfo {author} {\bibfnamefont {R.}~\bibnamefont {Kosloff}},\ }\bibfield  {title} {\bibinfo {title} {Quantum lubrication: Suppression of friction in a first-principles four-stroke heat engine},\ }\href {https://doi.org/10.1103/PhysRevE.73.025107} {\bibfield  {journal} {\bibinfo  {journal} {Phys. Rev. E}\ }\textbf {\bibinfo {volume} {73}},\ \bibinfo {pages} {025107} (\bibinfo {year} {2006})}\BibitemShut {NoStop}%
\bibitem [{\citenamefont {Rezek}\ and\ \citenamefont {Kosloff}(2006)}]{rezek2006}%
  \BibitemOpen
  \bibfield  {author} {\bibinfo {author} {\bibfnamefont {Y.}~\bibnamefont {Rezek}}\ and\ \bibinfo {author} {\bibfnamefont {R.}~\bibnamefont {Kosloff}},\ }\bibfield  {title} {\bibinfo {title} {Irreversible performance of a quantum harmonic heat engine},\ }\href {https://doi.org/10.1088/1367-2630/8/5/083} {\bibfield  {journal} {\bibinfo  {journal} {New J. Phys.}\ }\textbf {\bibinfo {volume} {8}},\ \bibinfo {pages} {83} (\bibinfo {year} {2006})}\BibitemShut {NoStop}%
\bibitem [{\citenamefont {Salamon}\ \emph {et~al.}(2009)\citenamefont {Salamon}, \citenamefont {Hoffmann}, \citenamefont {Rezek},\ and\ \citenamefont {Kosloff}}]{salamon2009}%
  \BibitemOpen
  \bibfield  {author} {\bibinfo {author} {\bibfnamefont {P.}~\bibnamefont {Salamon}}, \bibinfo {author} {\bibfnamefont {K.~H.}\ \bibnamefont {Hoffmann}}, \bibinfo {author} {\bibfnamefont {Y.}~\bibnamefont {Rezek}},\ and\ \bibinfo {author} {\bibfnamefont {R.}~\bibnamefont {Kosloff}},\ }\bibfield  {title} {\bibinfo {title} {Maximum work in minimum time from a conservative quantum system},\ }\href {https://doi.org/10.1039/B816102J} {\bibfield  {journal} {\bibinfo  {journal} {Phys. Chem. Chem. Phys.}\ }\textbf {\bibinfo {volume} {11}},\ \bibinfo {pages} {1027} (\bibinfo {year} {2009})}\BibitemShut {NoStop}%
\bibitem [{\citenamefont {Alecce}\ \emph {et~al.}(2015)\citenamefont {Alecce}, \citenamefont {Galve}, \citenamefont {Gullo}, \citenamefont {Dell’Anna}, \citenamefont {Plastina},\ and\ \citenamefont {Zambrini}}]{Alecce_2015}%
  \BibitemOpen
  \bibfield  {author} {\bibinfo {author} {\bibfnamefont {A.}~\bibnamefont {Alecce}}, \bibinfo {author} {\bibfnamefont {F.}~\bibnamefont {Galve}}, \bibinfo {author} {\bibfnamefont {N.~L.}\ \bibnamefont {Gullo}}, \bibinfo {author} {\bibfnamefont {L.}~\bibnamefont {Dell’Anna}}, \bibinfo {author} {\bibfnamefont {F.}~\bibnamefont {Plastina}},\ and\ \bibinfo {author} {\bibfnamefont {R.}~\bibnamefont {Zambrini}},\ }\bibfield  {title} {\bibinfo {title} {Quantum otto cycle with inner friction: finite-time and disorder effects},\ }\href {https://doi.org/10.1088/1367-2630/17/7/075007} {\bibfield  {journal} {\bibinfo  {journal} {New J. Phys.}\ }\textbf {\bibinfo {volume} {17}},\ \bibinfo {pages} {075007} (\bibinfo {year} {2015})}\BibitemShut {NoStop}%
\bibitem [{\citenamefont {Kosloff}(2013)}]{kosloff2013}%
  \BibitemOpen
  \bibfield  {author} {\bibinfo {author} {\bibfnamefont {R.}~\bibnamefont {Kosloff}},\ }\bibfield  {title} {\bibinfo {title} {Quantum thermodynamics: A dynamical viewpoint},\ }\href {https://doi.org/10.3390/e15062100} {\bibfield  {journal} {\bibinfo  {journal} {Entropy}\ }\textbf {\bibinfo {volume} {15}},\ \bibinfo {pages} {2100} (\bibinfo {year} {2013})}\BibitemShut {NoStop}%
\bibitem [{\citenamefont {Allahverdyan}\ and\ \citenamefont {Nieuwenhuizen}(2005)}]{PhysRevE.71.046107}%
  \BibitemOpen
  \bibfield  {author} {\bibinfo {author} {\bibfnamefont {A.~E.}\ \bibnamefont {Allahverdyan}}\ and\ \bibinfo {author} {\bibfnamefont {T.~M.}\ \bibnamefont {Nieuwenhuizen}},\ }\bibfield  {title} {\bibinfo {title} {Minimal work principle: Proof and counterexamples},\ }\href {https://doi.org/10.1103/PhysRevE.71.046107} {\bibfield  {journal} {\bibinfo  {journal} {Phys. Rev. E}\ }\textbf {\bibinfo {volume} {71}},\ \bibinfo {pages} {046107} (\bibinfo {year} {2005})}\BibitemShut {NoStop}%
\bibitem [{\citenamefont {Talkner}\ \emph {et~al.}(2007)\citenamefont {Talkner}, \citenamefont {Lutz},\ and\ \citenamefont {H\"anggi}}]{PhysRevE.75.050102}%
  \BibitemOpen
  \bibfield  {author} {\bibinfo {author} {\bibfnamefont {P.}~\bibnamefont {Talkner}}, \bibinfo {author} {\bibfnamefont {E.}~\bibnamefont {Lutz}},\ and\ \bibinfo {author} {\bibfnamefont {P.}~\bibnamefont {H\"anggi}},\ }\bibfield  {title} {\bibinfo {title} {Fluctuation theorems: Work is not an observable},\ }\href {https://doi.org/10.1103/PhysRevE.75.050102} {\bibfield  {journal} {\bibinfo  {journal} {Phys. Rev. E}\ }\textbf {\bibinfo {volume} {75}},\ \bibinfo {pages} {050102} (\bibinfo {year} {2007})}\BibitemShut {NoStop}%
\bibitem [{\citenamefont {Jarzynski}\ \emph {et~al.}(2015)\citenamefont {Jarzynski}, \citenamefont {Quan},\ and\ \citenamefont {Rahav}}]{PhysRevX.5.031038}%
  \BibitemOpen
  \bibfield  {author} {\bibinfo {author} {\bibfnamefont {C.}~\bibnamefont {Jarzynski}}, \bibinfo {author} {\bibfnamefont {H.~T.}\ \bibnamefont {Quan}},\ and\ \bibinfo {author} {\bibfnamefont {S.}~\bibnamefont {Rahav}},\ }\bibfield  {title} {\bibinfo {title} {Quantum-classical correspondence principle for work distributions},\ }\href {https://doi.org/10.1103/PhysRevX.5.031038} {\bibfield  {journal} {\bibinfo  {journal} {Phys. Rev. X}\ }\textbf {\bibinfo {volume} {5}},\ \bibinfo {pages} {031038} (\bibinfo {year} {2015})}\BibitemShut {NoStop}%
\bibitem [{\citenamefont {Plastina}\ \emph {et~al.}(2014)\citenamefont {Plastina}, \citenamefont {Alecce}, \citenamefont {Apollaro}, \citenamefont {Falcone}, \citenamefont {Francica}, \citenamefont {Galve}, \citenamefont {Lo~Gullo},\ and\ \citenamefont {Zambrini}}]{PhysRevLett.113.260601}%
  \BibitemOpen
  \bibfield  {author} {\bibinfo {author} {\bibfnamefont {F.}~\bibnamefont {Plastina}}, \bibinfo {author} {\bibfnamefont {A.}~\bibnamefont {Alecce}}, \bibinfo {author} {\bibfnamefont {T.~J.~G.}\ \bibnamefont {Apollaro}}, \bibinfo {author} {\bibfnamefont {G.}~\bibnamefont {Falcone}}, \bibinfo {author} {\bibfnamefont {G.}~\bibnamefont {Francica}}, \bibinfo {author} {\bibfnamefont {F.}~\bibnamefont {Galve}}, \bibinfo {author} {\bibfnamefont {N.}~\bibnamefont {Lo~Gullo}},\ and\ \bibinfo {author} {\bibfnamefont {R.}~\bibnamefont {Zambrini}},\ }\bibfield  {title} {\bibinfo {title} {Irreversible work and inner friction in quantum thermodynamic processes},\ }\href {https://doi.org/10.1103/PhysRevLett.113.260601} {\bibfield  {journal} {\bibinfo  {journal} {Phys. Rev. Lett.}\ }\textbf {\bibinfo {volume} {113}},\ \bibinfo {pages} {260601} (\bibinfo {year} {2014})}\BibitemShut {NoStop}%
\bibitem [{\citenamefont {Feldmann}\ and\ \citenamefont {Kosloff}(2012)}]{feldmann2012}%
  \BibitemOpen
  \bibfield  {author} {\bibinfo {author} {\bibfnamefont {T.}~\bibnamefont {Feldmann}}\ and\ \bibinfo {author} {\bibfnamefont {R.}~\bibnamefont {Kosloff}},\ }\bibfield  {title} {\bibinfo {title} {Short time cycles of purely quantum refrigerators},\ }\href {https://doi.org/10.1103/PhysRevE.85.051114} {\bibfield  {journal} {\bibinfo  {journal} {Phys. Rev. E}\ }\textbf {\bibinfo {volume} {85}},\ \bibinfo {pages} {051114} (\bibinfo {year} {2012})}\BibitemShut {NoStop}%
\bibitem [{\citenamefont {Balian}(1989)}]{balian1989}%
  \BibitemOpen
  \bibfield  {author} {\bibinfo {author} {\bibfnamefont {R.}~\bibnamefont {Balian}},\ }\bibfield  {title} {\bibinfo {title} {Gain of information in a quantum measurement},\ }\href {https://doi.org/10.1088/0143-0807/10/3/010} {\bibfield  {journal} {\bibinfo  {journal} {Eur. J. Phys.}\ }\textbf {\bibinfo {volume} {10}},\ \bibinfo {pages} {208} (\bibinfo {year} {1989})}\BibitemShut {NoStop}%
\bibitem [{\citenamefont {Polkovnikov}(2011)}]{POLKOVNIKOV2011486}%
  \BibitemOpen
  \bibfield  {author} {\bibinfo {author} {\bibfnamefont {A.}~\bibnamefont {Polkovnikov}},\ }\bibfield  {title} {\bibinfo {title} {Microscopic diagonal entropy and its connection to basic thermodynamic relations},\ }\href {https://doi.org/https://doi.org/10.1016/j.aop.2010.08.004} {\bibfield  {journal} {\bibinfo  {journal} {Ann. Phys}\ }\textbf {\bibinfo {volume} {326}},\ \bibinfo {pages} {486} (\bibinfo {year} {2011})}\BibitemShut {NoStop}%
\bibitem [{\citenamefont {Santos}\ \emph {et~al.}(2011)\citenamefont {Santos}, \citenamefont {Polkovnikov},\ and\ \citenamefont {Rigol}}]{PhysRevLett.107.040601}%
  \BibitemOpen
  \bibfield  {author} {\bibinfo {author} {\bibfnamefont {L.~F.}\ \bibnamefont {Santos}}, \bibinfo {author} {\bibfnamefont {A.}~\bibnamefont {Polkovnikov}},\ and\ \bibinfo {author} {\bibfnamefont {M.}~\bibnamefont {Rigol}},\ }\bibfield  {title} {\bibinfo {title} {Entropy of isolated quantum systems after a quench},\ }\href {https://doi.org/10.1103/PhysRevLett.107.040601} {\bibfield  {journal} {\bibinfo  {journal} {Phys. Rev. Lett.}\ }\textbf {\bibinfo {volume} {107}},\ \bibinfo {pages} {040601} (\bibinfo {year} {2011})}\BibitemShut {NoStop}%
\bibitem [{\citenamefont {Ikeda}\ \emph {et~al.}(2015)\citenamefont {Ikeda}, \citenamefont {Sakumichi}, \citenamefont {Polkovnikov},\ and\ \citenamefont {Ueda}}]{IKEDA2015338}%
  \BibitemOpen
  \bibfield  {author} {\bibinfo {author} {\bibfnamefont {T.~N.}\ \bibnamefont {Ikeda}}, \bibinfo {author} {\bibfnamefont {N.}~\bibnamefont {Sakumichi}}, \bibinfo {author} {\bibfnamefont {A.}~\bibnamefont {Polkovnikov}},\ and\ \bibinfo {author} {\bibfnamefont {M.}~\bibnamefont {Ueda}},\ }\bibfield  {title} {\bibinfo {title} {The second law of thermodynamics under unitary evolution and external operations},\ }\href {https://doi.org/https://doi.org/10.1016/j.aop.2015.01.003} {\bibfield  {journal} {\bibinfo  {journal} {Ann. Phys.}\ }\textbf {\bibinfo {volume} {354}},\ \bibinfo {pages} {338} (\bibinfo {year} {2015})}\BibitemShut {NoStop}%
\bibitem [{\citenamefont {Kosloff}\ and\ \citenamefont {Rezek}(2017)}]{e19040136}%
  \BibitemOpen
  \bibfield  {author} {\bibinfo {author} {\bibfnamefont {R.}~\bibnamefont {Kosloff}}\ and\ \bibinfo {author} {\bibfnamefont {Y.}~\bibnamefont {Rezek}},\ }\bibfield  {title} {\bibinfo {title} {The quantum harmonic {Otto} cycle},\ }\href {https://www.mdpi.com/1099-4300/19/4/136} {\bibfield  {journal} {\bibinfo  {journal} {Entropy}\ }\textbf {\bibinfo {volume} {19}} (\bibinfo {year} {2017})}\BibitemShut {NoStop}%
\bibitem [{\citenamefont {Francica}\ \emph {et~al.}(2019)\citenamefont {Francica}, \citenamefont {Goold},\ and\ \citenamefont {Plastina}}]{PhysRevE.99.042105}%
  \BibitemOpen
  \bibfield  {author} {\bibinfo {author} {\bibfnamefont {G.}~\bibnamefont {Francica}}, \bibinfo {author} {\bibfnamefont {J.}~\bibnamefont {Goold}},\ and\ \bibinfo {author} {\bibfnamefont {F.}~\bibnamefont {Plastina}},\ }\bibfield  {title} {\bibinfo {title} {Role of coherence in the nonequilibrium thermodynamics of quantum systems},\ }\href {https://doi.org/10.1103/PhysRevE.99.042105} {\bibfield  {journal} {\bibinfo  {journal} {Phys. Rev. E}\ }\textbf {\bibinfo {volume} {99}},\ \bibinfo {pages} {042105} (\bibinfo {year} {2019})}\BibitemShut {NoStop}%
\bibitem [{\citenamefont {Baumgratz}\ \emph {et~al.}(2014)\citenamefont {Baumgratz}, \citenamefont {Cramer},\ and\ \citenamefont {Plenio}}]{PhysRevLett.113.140401}%
  \BibitemOpen
  \bibfield  {author} {\bibinfo {author} {\bibfnamefont {T.}~\bibnamefont {Baumgratz}}, \bibinfo {author} {\bibfnamefont {M.}~\bibnamefont {Cramer}},\ and\ \bibinfo {author} {\bibfnamefont {M.~B.}\ \bibnamefont {Plenio}},\ }\bibfield  {title} {\bibinfo {title} {Quantifying coherence},\ }\href {https://doi.org/10.1103/PhysRevLett.113.140401} {\bibfield  {journal} {\bibinfo  {journal} {Phys. Rev. Lett.}\ }\textbf {\bibinfo {volume} {113}},\ \bibinfo {pages} {140401} (\bibinfo {year} {2014})}\BibitemShut {NoStop}%
\bibitem [{\citenamefont {Streltsov}\ \emph {et~al.}(2017)\citenamefont {Streltsov}, \citenamefont {Adesso},\ and\ \citenamefont {Plenio}}]{RevModPhys.89.041003}%
  \BibitemOpen
  \bibfield  {author} {\bibinfo {author} {\bibfnamefont {A.}~\bibnamefont {Streltsov}}, \bibinfo {author} {\bibfnamefont {G.}~\bibnamefont {Adesso}},\ and\ \bibinfo {author} {\bibfnamefont {M.~B.}\ \bibnamefont {Plenio}},\ }\bibfield  {title} {\bibinfo {title} {Colloquium: Quantum coherence as a resource},\ }\href {https://doi.org/10.1103/RevModPhys.89.041003} {\bibfield  {journal} {\bibinfo  {journal} {Rev. Mod. Phys.}\ }\textbf {\bibinfo {volume} {89}},\ \bibinfo {pages} {041003} (\bibinfo {year} {2017})}\BibitemShut {NoStop}%
\bibitem [{\citenamefont {D'Alessio}\ \emph {et~al.}(2016{\natexlab{a}})\citenamefont {D'Alessio}, \citenamefont {Kafri}, \citenamefont {Polkovnikov},\ and\ \citenamefont {Rigol}}]{dalessio2016}%
  \BibitemOpen
  \bibfield  {author} {\bibinfo {author} {\bibfnamefont {L.}~\bibnamefont {D'Alessio}}, \bibinfo {author} {\bibfnamefont {Y.}~\bibnamefont {Kafri}}, \bibinfo {author} {\bibfnamefont {A.}~\bibnamefont {Polkovnikov}},\ and\ \bibinfo {author} {\bibfnamefont {M.}~\bibnamefont {Rigol}},\ }\bibfield  {title} {\bibinfo {title} {From quantum chaos and eigenstate thermalization to statistical mechanics and thermodynamics},\ }\href {https://doi.org/https://doi.org/10.1080/00018732.2016.1198134} {\bibfield  {journal} {\bibinfo  {journal} {Adv. Phys.}\ }\textbf {\bibinfo {volume} {65}},\ \bibinfo {pages} {239} (\bibinfo {year} {2016}{\natexlab{a}})}\BibitemShut {NoStop}%
\bibitem [{\citenamefont {Mori}\ \emph {et~al.}(2018)\citenamefont {Mori}, \citenamefont {Ikeda}, \citenamefont {Kaminishi},\ and\ \citenamefont {Ueda}}]{mori2018}%
  \BibitemOpen
  \bibfield  {author} {\bibinfo {author} {\bibfnamefont {T.}~\bibnamefont {Mori}}, \bibinfo {author} {\bibfnamefont {T.~N.}\ \bibnamefont {Ikeda}}, \bibinfo {author} {\bibfnamefont {E.}~\bibnamefont {Kaminishi}},\ and\ \bibinfo {author} {\bibfnamefont {M.}~\bibnamefont {Ueda}},\ }\bibfield  {title} {\bibinfo {title} {Thermalization and prethermalization in isolated quantum systems: a theoretical overview},\ }\href@noop {} {\bibfield  {journal} {\bibinfo  {journal} {J. Phys. B: At. Mol. Opt. Phys.}\ }\textbf {\bibinfo {volume} {51}},\ \bibinfo {pages} {112001} (\bibinfo {year} {2018})}\BibitemShut {NoStop}%
\bibitem [{\citenamefont {Haake}\ \emph {et~al.}(2018)\citenamefont {Haake}, \citenamefont {Gnutzmann},\ and\ \citenamefont {Ku{\'s}}}]{haake2018}%
  \BibitemOpen
  \bibfield  {author} {\bibinfo {author} {\bibfnamefont {F.}~\bibnamefont {Haake}}, \bibinfo {author} {\bibfnamefont {S.}~\bibnamefont {Gnutzmann}},\ and\ \bibinfo {author} {\bibfnamefont {M.}~\bibnamefont {Ku{\'s}}},\ }\href {https://doi.org/10.1007/978-3-319-97580-1} {\emph {\bibinfo {title} {Quantum Signatures of Chaos}}},\ \bibinfo {edition} {4th}\ ed.,\ Springer Series in Synergetics\ (\bibinfo  {publisher} {Springer},\ \bibinfo {year} {2018})\BibitemShut {NoStop}%
\bibitem [{\citenamefont {Pfeuty}(1970)}]{PFEUTY197079}%
  \BibitemOpen
  \bibfield  {author} {\bibinfo {author} {\bibfnamefont {P.}~\bibnamefont {Pfeuty}},\ }\bibfield  {title} {\bibinfo {title} {The one-dimensional {Ising} model with a transverse field},\ }\href {https://doi.org/https://doi.org/10.1016/0003-4916(70)90270-8} {\bibfield  {journal} {\bibinfo  {journal} {Ann. Phys.}\ }\textbf {\bibinfo {volume} {57}},\ \bibinfo {pages} {79} (\bibinfo {year} {1970})}\BibitemShut {NoStop}%
\bibitem [{\citenamefont {Mbeng}\ \emph {et~al.}(2024)\citenamefont {Mbeng}, \citenamefont {Russomanno},\ and\ \citenamefont {Santoro}}]{mbeng2020quantum}%
  \BibitemOpen
  \bibfield  {author} {\bibinfo {author} {\bibfnamefont {G.~B.}\ \bibnamefont {Mbeng}}, \bibinfo {author} {\bibfnamefont {A.}~\bibnamefont {Russomanno}},\ and\ \bibinfo {author} {\bibfnamefont {G.~E.}\ \bibnamefont {Santoro}},\ }\bibfield  {title} {\bibinfo {title} {{The quantum {Ising} chain for beginners}},\ }\href {https://doi.org/10.21468/SciPostPhysLectNotes.82} {\bibfield  {journal} {\bibinfo  {journal} {SciPost Phys. Lect. Notes}\ ,\ \bibinfo {pages} {82}} (\bibinfo {year} {2024})}\BibitemShut {NoStop}%
\bibitem [{\citenamefont {Sachdev}(2011)}]{Sachdev_2011}%
  \BibitemOpen
  \bibfield  {author} {\bibinfo {author} {\bibfnamefont {S.}~\bibnamefont {Sachdev}},\ }\href@noop {} {\emph {\bibinfo {title} {Quantum Phase Transitions}}},\ \bibinfo {edition} {2nd}\ ed.\ (\bibinfo  {publisher} {Cambridge University Press},\ \bibinfo {year} {2011})\BibitemShut {NoStop}%
\bibitem [{\citenamefont {Kim}\ and\ \citenamefont {Huse}(2013)}]{kim2013}%
  \BibitemOpen
  \bibfield  {author} {\bibinfo {author} {\bibfnamefont {H.}~\bibnamefont {Kim}}\ and\ \bibinfo {author} {\bibfnamefont {D.~A.}\ \bibnamefont {Huse}},\ }\bibfield  {title} {\bibinfo {title} {Ballistic spreading of entanglement in a diffusive nonintegrable system},\ }\href {https://doi.org/10.1103/PhysRevLett.111.127205} {\bibfield  {journal} {\bibinfo  {journal} {Phys. Rev. Lett.}\ }\textbf {\bibinfo {volume} {111}},\ \bibinfo {pages} {127205} (\bibinfo {year} {2013})}\BibitemShut {NoStop}%
\bibitem [{\citenamefont {Kim}\ \emph {et~al.}(2014)\citenamefont {Kim}, \citenamefont {Ikeda},\ and\ \citenamefont {Huse}}]{kim2014}%
  \BibitemOpen
  \bibfield  {author} {\bibinfo {author} {\bibfnamefont {H.}~\bibnamefont {Kim}}, \bibinfo {author} {\bibfnamefont {T.~N.}\ \bibnamefont {Ikeda}},\ and\ \bibinfo {author} {\bibfnamefont {D.~A.}\ \bibnamefont {Huse}},\ }\bibfield  {title} {\bibinfo {title} {Testing whether all eigenstates obey the eigenstate thermalization hypothesis},\ }\href {https://doi.org/10.1103/PhysRevE.90.052105} {\bibfield  {journal} {\bibinfo  {journal} {Phys. Rev. E}\ }\textbf {\bibinfo {volume} {90}},\ \bibinfo {pages} {052105} (\bibinfo {year} {2014})}\BibitemShut {NoStop}%
\bibitem [{\citenamefont {Britton}\ \emph {et~al.}(2012)\citenamefont {Britton}, \citenamefont {Sawyer}, \citenamefont {Keith}, \citenamefont {Wang}, \citenamefont {Freericks}, \citenamefont {Uys}, \citenamefont {Biercuk},\ and\ \citenamefont {Bollinger}}]{britton2012}%
  \BibitemOpen
  \bibfield  {author} {\bibinfo {author} {\bibfnamefont {J.~W.}\ \bibnamefont {Britton}}, \bibinfo {author} {\bibfnamefont {B.~C.}\ \bibnamefont {Sawyer}}, \bibinfo {author} {\bibfnamefont {A.~C.}\ \bibnamefont {Keith}}, \bibinfo {author} {\bibfnamefont {C.-C.~J.}\ \bibnamefont {Wang}}, \bibinfo {author} {\bibfnamefont {J.~K.}\ \bibnamefont {Freericks}}, \bibinfo {author} {\bibfnamefont {H.}~\bibnamefont {Uys}}, \bibinfo {author} {\bibfnamefont {M.~J.}\ \bibnamefont {Biercuk}},\ and\ \bibinfo {author} {\bibfnamefont {J.~J.}\ \bibnamefont {Bollinger}},\ }\bibfield  {title} {\bibinfo {title} {Engineered two-dimensional {Ising} interactions in a trapped-ion quantum simulator with hundreds of spins},\ }\href {https://doi.org/10.1038/nature10981} {\bibfield  {journal} {\bibinfo  {journal} {Nature}\ }\textbf {\bibinfo {volume} {484}},\ \bibinfo {pages} {489} (\bibinfo {year} {2012})}\BibitemShut {NoStop}%
\bibitem [{\citenamefont {Monroe}\ \emph {et~al.}(2021)\citenamefont {Monroe}, \citenamefont {Campbell}, \citenamefont {Duan}, \citenamefont {Gong}, \citenamefont {Gorshkov}, \citenamefont {Hess}, \citenamefont {Islam}, \citenamefont {Kim}, \citenamefont {Linke}, \citenamefont {Pagano}, \citenamefont {Richerme}, \citenamefont {Senko},\ and\ \citenamefont {Yao}}]{RevModPhys.93.025001}%
  \BibitemOpen
  \bibfield  {author} {\bibinfo {author} {\bibfnamefont {C.}~\bibnamefont {Monroe}}, \bibinfo {author} {\bibfnamefont {W.~C.}\ \bibnamefont {Campbell}}, \bibinfo {author} {\bibfnamefont {L.-M.}\ \bibnamefont {Duan}}, \bibinfo {author} {\bibfnamefont {Z.-X.}\ \bibnamefont {Gong}}, \bibinfo {author} {\bibfnamefont {A.~V.}\ \bibnamefont {Gorshkov}}, \bibinfo {author} {\bibfnamefont {P.~W.}\ \bibnamefont {Hess}}, \bibinfo {author} {\bibfnamefont {R.}~\bibnamefont {Islam}}, \bibinfo {author} {\bibfnamefont {K.}~\bibnamefont {Kim}}, \bibinfo {author} {\bibfnamefont {N.~M.}\ \bibnamefont {Linke}}, \bibinfo {author} {\bibfnamefont {G.}~\bibnamefont {Pagano}}, \bibinfo {author} {\bibfnamefont {P.}~\bibnamefont {Richerme}}, \bibinfo {author} {\bibfnamefont {C.}~\bibnamefont {Senko}},\ and\ \bibinfo {author} {\bibfnamefont {N.~Y.}\ \bibnamefont {Yao}},\ }\bibfield  {title} {\bibinfo {title} {Programmable quantum simulations of spin systems with trapped ions},\ }\href {https://doi.org/10.1103/RevModPhys.93.025001}
  {\bibfield  {journal} {\bibinfo  {journal} {Rev. Mod. Phys.}\ }\textbf {\bibinfo {volume} {93}},\ \bibinfo {pages} {025001} (\bibinfo {year} {2021})}\BibitemShut {NoStop}%
\bibitem [{\citenamefont {Labuhn}\ \emph {et~al.}(2016)\citenamefont {Labuhn}, \citenamefont {Barredo}, \citenamefont {Ravets}, \citenamefont {de~L{\'e}s{\'e}leuc}, \citenamefont {Macr{\`i}}, \citenamefont {Lahaye},\ and\ \citenamefont {Browaeys}}]{Labuhn2016}%
  \BibitemOpen
  \bibfield  {author} {\bibinfo {author} {\bibfnamefont {H.}~\bibnamefont {Labuhn}}, \bibinfo {author} {\bibfnamefont {D.}~\bibnamefont {Barredo}}, \bibinfo {author} {\bibfnamefont {S.}~\bibnamefont {Ravets}}, \bibinfo {author} {\bibfnamefont {S.}~\bibnamefont {de~L{\'e}s{\'e}leuc}}, \bibinfo {author} {\bibfnamefont {T.}~\bibnamefont {Macr{\`i}}}, \bibinfo {author} {\bibfnamefont {T.}~\bibnamefont {Lahaye}},\ and\ \bibinfo {author} {\bibfnamefont {A.}~\bibnamefont {Browaeys}},\ }\bibfield  {title} {\bibinfo {title} {Tunable two-dimensional arrays of single {R}ydberg atoms for realizing quantum {I}sing models},\ }\href {https://doi.org/10.1038/nature18274} {\bibfield  {journal} {\bibinfo  {journal} {Nature}\ }\textbf {\bibinfo {volume} {534}},\ \bibinfo {pages} {667} (\bibinfo {year} {2016})}\BibitemShut {NoStop}%
\bibitem [{\citenamefont {Browaeys}\ and\ \citenamefont {Lahaye}(2020)}]{browaeys2020}%
  \BibitemOpen
  \bibfield  {author} {\bibinfo {author} {\bibfnamefont {A.}~\bibnamefont {Browaeys}}\ and\ \bibinfo {author} {\bibfnamefont {T.}~\bibnamefont {Lahaye}},\ }\bibfield  {title} {\bibinfo {title} {Many-body physics with individually controlled rydberg atoms},\ }\href {https://doi.org/10.1038/s41567-019-0733-z} {\bibfield  {journal} {\bibinfo  {journal} {Nat. Phys.}\ }\textbf {\bibinfo {volume} {16}},\ \bibinfo {pages} {132} (\bibinfo {year} {2020})}\BibitemShut {NoStop}%
\bibitem [{\citenamefont {Piccitto}\ \emph {et~al.}(2022)\citenamefont {Piccitto}, \citenamefont {Campisi},\ and\ \citenamefont {Rossini}}]{piccitto2022}%
  \BibitemOpen
  \bibfield  {author} {\bibinfo {author} {\bibfnamefont {G.}~\bibnamefont {Piccitto}}, \bibinfo {author} {\bibfnamefont {M.}~\bibnamefont {Campisi}},\ and\ \bibinfo {author} {\bibfnamefont {D.}~\bibnamefont {Rossini}},\ }\bibfield  {title} {\bibinfo {title} {The {Ising} critical quantum {Otto} engine},\ }\href {https://doi.org/10.1088/1367-2630/ac963b} {\bibfield  {journal} {\bibinfo  {journal} {New J. Phys.}\ }\textbf {\bibinfo {volume} {24}},\ \bibinfo {pages} {103023} (\bibinfo {year} {2022})}\BibitemShut {NoStop}%
\bibitem [{\citenamefont {B.~S}\ \emph {et~al.}(2020)\citenamefont {B.~S}, \citenamefont {Mukherjee}, \citenamefont {Divakaran},\ and\ \citenamefont {del Campo}}]{revathy2020}%
  \BibitemOpen
  \bibfield  {author} {\bibinfo {author} {\bibfnamefont {R.}~\bibnamefont {B.~S}}, \bibinfo {author} {\bibfnamefont {V.}~\bibnamefont {Mukherjee}}, \bibinfo {author} {\bibfnamefont {U.}~\bibnamefont {Divakaran}},\ and\ \bibinfo {author} {\bibfnamefont {A.}~\bibnamefont {del Campo}},\ }\bibfield  {title} {\bibinfo {title} {Universal finite-time thermodynamics of many-body quantum machines from {Kibble}-{Zurek} scaling},\ }\href {https://doi.org/10.1103/PhysRevResearch.2.043247} {\bibfield  {journal} {\bibinfo  {journal} {Phys. Rev. Res.}\ }\textbf {\bibinfo {volume} {2}},\ \bibinfo {pages} {043247} (\bibinfo {year} {2020})}\BibitemShut {NoStop}%
\bibitem [{\citenamefont {Williamson}\ and\ \citenamefont {Davis}(2024)}]{PhysRevB.109.024310}%
  \BibitemOpen
  \bibfield  {author} {\bibinfo {author} {\bibfnamefont {L.~A.}\ \bibnamefont {Williamson}}\ and\ \bibinfo {author} {\bibfnamefont {M.~J.}\ \bibnamefont {Davis}},\ }\bibfield  {title} {\bibinfo {title} {Many-body enhancement in a spin-chain quantum heat engine},\ }\href {https://doi.org/10.1103/PhysRevB.109.024310} {\bibfield  {journal} {\bibinfo  {journal} {Phys. Rev. B}\ }\textbf {\bibinfo {volume} {109}},\ \bibinfo {pages} {024310} (\bibinfo {year} {2024})}\BibitemShut {NoStop}%
\bibitem [{\citenamefont {Arezzo}\ \emph {et~al.}(2024)\citenamefont {Arezzo}, \citenamefont {Rossini},\ and\ \citenamefont {Piccitto}}]{PhysRevB.109.224309}%
  \BibitemOpen
  \bibfield  {author} {\bibinfo {author} {\bibfnamefont {V.~R.}\ \bibnamefont {Arezzo}}, \bibinfo {author} {\bibfnamefont {D.}~\bibnamefont {Rossini}},\ and\ \bibinfo {author} {\bibfnamefont {G.}~\bibnamefont {Piccitto}},\ }\bibfield  {title} {\bibinfo {title} {Many-body quantum heat engines based on free fermion systems},\ }\href {https://doi.org/10.1103/PhysRevB.109.224309} {\bibfield  {journal} {\bibinfo  {journal} {Phys. Rev. B}\ }\textbf {\bibinfo {volume} {109}},\ \bibinfo {pages} {224309} (\bibinfo {year} {2024})}\BibitemShut {NoStop}%
\bibitem [{\citenamefont {Sajitha}\ \emph {et~al.}(2025)\citenamefont {Sajitha}, \citenamefont {Santra}, \citenamefont {Davis},\ and\ \citenamefont {Williamson}}]{jthm-7c2j}%
  \BibitemOpen
  \bibfield  {author} {\bibinfo {author} {\bibfnamefont {V.~M.}\ \bibnamefont {Sajitha}}, \bibinfo {author} {\bibfnamefont {B.}~\bibnamefont {Santra}}, \bibinfo {author} {\bibfnamefont {M.~J.}\ \bibnamefont {Davis}},\ and\ \bibinfo {author} {\bibfnamefont {L.~A.}\ \bibnamefont {Williamson}},\ }\bibfield  {title} {\bibinfo {title} {Quantum thermal machine regimes in the transverse-field {I}sing model},\ }\href {https://doi.org/10.1103/jthm-7c2j} {\bibfield  {journal} {\bibinfo  {journal} {Phys. Rev. A}\ }\textbf {\bibinfo {volume} {111}},\ \bibinfo {pages} {062213} (\bibinfo {year} {2025})}\BibitemShut {NoStop}%
\bibitem [{\citenamefont {Wang}(2020)}]{wang2020}%
  \BibitemOpen
  \bibfield  {author} {\bibinfo {author} {\bibfnamefont {Q.}~\bibnamefont {Wang}},\ }\bibfield  {title} {\bibinfo {title} {Performance of quantum heat engines under the influence of long-range interactions},\ }\href {https://doi.org/10.1103/PhysRevE.102.012138} {\bibfield  {journal} {\bibinfo  {journal} {Phys. Rev. E}\ }\textbf {\bibinfo {volume} {102}},\ \bibinfo {pages} {012138} (\bibinfo {year} {2020})}\BibitemShut {NoStop}%
\bibitem [{\citenamefont {Solfanelli}\ \emph {et~al.}(2023)\citenamefont {Solfanelli}, \citenamefont {Giachetti}, \citenamefont {Campisi}, \citenamefont {Ruffo},\ and\ \citenamefont {Defenu}}]{Solfanelli_2023}%
  \BibitemOpen
  \bibfield  {author} {\bibinfo {author} {\bibfnamefont {A.}~\bibnamefont {Solfanelli}}, \bibinfo {author} {\bibfnamefont {G.}~\bibnamefont {Giachetti}}, \bibinfo {author} {\bibfnamefont {M.}~\bibnamefont {Campisi}}, \bibinfo {author} {\bibfnamefont {S.}~\bibnamefont {Ruffo}},\ and\ \bibinfo {author} {\bibfnamefont {N.}~\bibnamefont {Defenu}},\ }\bibfield  {title} {\bibinfo {title} {Quantum heat engine with long-range advantages},\ }\href {https://doi.org/10.1088/1367-2630/acc04e} {\bibfield  {journal} {\bibinfo  {journal} {New J. Phys.}\ }\textbf {\bibinfo {volume} {25}},\ \bibinfo {pages} {033030} (\bibinfo {year} {2023})}\BibitemShut {NoStop}%
\bibitem [{\citenamefont {Uzelac}\ \emph {et~al.}(1980)\citenamefont {Uzelac}, \citenamefont {Jullien},\ and\ \citenamefont {Pfeuty}}]{PhysRevB.22.436}%
  \BibitemOpen
  \bibfield  {author} {\bibinfo {author} {\bibfnamefont {K.}~\bibnamefont {Uzelac}}, \bibinfo {author} {\bibfnamefont {R.}~\bibnamefont {Jullien}},\ and\ \bibinfo {author} {\bibfnamefont {P.}~\bibnamefont {Pfeuty}},\ }\bibfield  {title} {\bibinfo {title} {One-dimensional transverse-field ising model in a complex longitudinal field from a real-space renormalization-group method at $t=0$},\ }\href {https://doi.org/10.1103/PhysRevB.22.436} {\bibfield  {journal} {\bibinfo  {journal} {Phys. Rev. B}\ }\textbf {\bibinfo {volume} {22}},\ \bibinfo {pages} {436} (\bibinfo {year} {1980})}\BibitemShut {NoStop}%
\bibitem [{\citenamefont {Sinitsyn}\ and\ \citenamefont {Li}(2016)}]{sinitsyn2016}%
  \BibitemOpen
  \bibfield  {author} {\bibinfo {author} {\bibfnamefont {N.~A.}\ \bibnamefont {Sinitsyn}}\ and\ \bibinfo {author} {\bibfnamefont {F.}~\bibnamefont {Li}},\ }\bibfield  {title} {\bibinfo {title} {Solvable multistate model of {Landau}-{Zener} transitions in cavity {QED}},\ }\href {https://doi.org/10.1103/PhysRevA.93.063859} {\bibfield  {journal} {\bibinfo  {journal} {Phys. Rev. A}\ }\textbf {\bibinfo {volume} {93}},\ \bibinfo {pages} {063859} (\bibinfo {year} {2016})}\BibitemShut {NoStop}%
\bibitem [{\citenamefont {Gaveau}\ and\ \citenamefont {Schulman}(1997)}]{GAVEAU1997347}%
  \BibitemOpen
  \bibfield  {author} {\bibinfo {author} {\bibfnamefont {B.}~\bibnamefont {Gaveau}}\ and\ \bibinfo {author} {\bibfnamefont {L.}~\bibnamefont {Schulman}},\ }\bibfield  {title} {\bibinfo {title} {A general framework for non-equilibrium phenomena: the master equation and its formal consequences},\ }\href {https://doi.org/https://doi.org/10.1016/S0375-9601(97)00185-0} {\bibfield  {journal} {\bibinfo  {journal} {Phys. Lett. A}\ }\textbf {\bibinfo {volume} {229}},\ \bibinfo {pages} {347} (\bibinfo {year} {1997})}\BibitemShut {NoStop}%
\bibitem [{\citenamefont {Skrzypczyk}\ \emph {et~al.}(2014)\citenamefont {Skrzypczyk}, \citenamefont {Short},\ and\ \citenamefont {Popescu}}]{Skrzypczyk2014}%
  \BibitemOpen
  \bibfield  {author} {\bibinfo {author} {\bibfnamefont {P.}~\bibnamefont {Skrzypczyk}}, \bibinfo {author} {\bibfnamefont {A.~J.}\ \bibnamefont {Short}},\ and\ \bibinfo {author} {\bibfnamefont {S.}~\bibnamefont {Popescu}},\ }\bibfield  {title} {\bibinfo {title} {Work extraction and thermodynamics for individual quantum systems},\ }\href {https://doi.org/10.1038/ncomms5185} {\bibfield  {journal} {\bibinfo  {journal} {Nat. Commun}\ }\textbf {\bibinfo {volume} {5}},\ \bibinfo {pages} {4185} (\bibinfo {year} {2014})}\BibitemShut {NoStop}%
\bibitem [{\citenamefont {Parrondo}\ \emph {et~al.}(2015)\citenamefont {Parrondo}, \citenamefont {Horowitz},\ and\ \citenamefont {Sagawa}}]{Parrondo2015}%
  \BibitemOpen
  \bibfield  {author} {\bibinfo {author} {\bibfnamefont {J.~M.~R.}\ \bibnamefont {Parrondo}}, \bibinfo {author} {\bibfnamefont {J.~M.}\ \bibnamefont {Horowitz}},\ and\ \bibinfo {author} {\bibfnamefont {T.}~\bibnamefont {Sagawa}},\ }\bibfield  {title} {\bibinfo {title} {Thermodynamics of information},\ }\href {https://doi.org/10.1038/nphys3230} {\bibfield  {journal} {\bibinfo  {journal} {Nat. Phys.}\ }\textbf {\bibinfo {volume} {11}},\ \bibinfo {pages} {131} (\bibinfo {year} {2015})}\BibitemShut {NoStop}%
\bibitem [{\citenamefont {de~Oliveira~Junior}\ \emph {et~al.}(2025)\citenamefont {de~Oliveira~Junior}, \citenamefont {Brask},\ and\ \citenamefont {Lipka-Bartosik}}]{PhysRevLett.134.050401}%
  \BibitemOpen
  \bibfield  {author} {\bibinfo {author} {\bibfnamefont {A.}~\bibnamefont {de~Oliveira~Junior}}, \bibinfo {author} {\bibfnamefont {J.~B.}\ \bibnamefont {Brask}},\ and\ \bibinfo {author} {\bibfnamefont {P.}~\bibnamefont {Lipka-Bartosik}},\ }\bibfield  {title} {\bibinfo {title} {Heat as a witness of quantum properties},\ }\href {https://doi.org/10.1103/PhysRevLett.134.050401} {\bibfield  {journal} {\bibinfo  {journal} {Phys. Rev. Lett.}\ }\textbf {\bibinfo {volume} {134}},\ \bibinfo {pages} {050401} (\bibinfo {year} {2025})}\BibitemShut {NoStop}%
\bibitem [{\citenamefont {Dziarmaga}(2005)}]{dziarmaga2005}%
  \BibitemOpen
  \bibfield  {author} {\bibinfo {author} {\bibfnamefont {J.}~\bibnamefont {Dziarmaga}},\ }\bibfield  {title} {\bibinfo {title} {Dynamics of a quantum phase transition: Exact solution of the quantum {Ising} model},\ }\href {https://doi.org/10.1103/PhysRevLett.95.245701} {\bibfield  {journal} {\bibinfo  {journal} {Phys. Rev. Lett.}\ }\textbf {\bibinfo {volume} {95}},\ \bibinfo {pages} {245701} (\bibinfo {year} {2005})}\BibitemShut {NoStop}%
\bibitem [{\citenamefont {Quan}\ \emph {et~al.}(2007)\citenamefont {Quan}, \citenamefont {Liu}, \citenamefont {Sun},\ and\ \citenamefont {Nori}}]{quan2007}%
  \BibitemOpen
  \bibfield  {author} {\bibinfo {author} {\bibfnamefont {H.~T.}\ \bibnamefont {Quan}}, \bibinfo {author} {\bibfnamefont {Y.-X.}\ \bibnamefont {Liu}}, \bibinfo {author} {\bibfnamefont {C.~P.}\ \bibnamefont {Sun}},\ and\ \bibinfo {author} {\bibfnamefont {F.}~\bibnamefont {Nori}},\ }\bibfield  {title} {\bibinfo {title} {Quantum thermodynamic cycles and quantum heat engines},\ }\href {https://doi.org/10.1103/PhysRevE.76.031105} {\bibfield  {journal} {\bibinfo  {journal} {Phys. Rev. E}\ }\textbf {\bibinfo {volume} {76}},\ \bibinfo {pages} {031105} (\bibinfo {year} {2007})}\BibitemShut {NoStop}%
\bibitem [{\citenamefont {Rigol}\ \emph {et~al.}(2007)\citenamefont {Rigol}, \citenamefont {Dunjko}, \citenamefont {Yurovsky},\ and\ \citenamefont {Olshanii}}]{rigol2007}%
  \BibitemOpen
  \bibfield  {author} {\bibinfo {author} {\bibfnamefont {M.}~\bibnamefont {Rigol}}, \bibinfo {author} {\bibfnamefont {V.}~\bibnamefont {Dunjko}}, \bibinfo {author} {\bibfnamefont {V.}~\bibnamefont {Yurovsky}},\ and\ \bibinfo {author} {\bibfnamefont {M.}~\bibnamefont {Olshanii}},\ }\bibfield  {title} {\bibinfo {title} {Relaxation in a completely integrable many-body quantum system: An ab initio study of the dynamics of the highly excited states of 1{D} lattice hard-core bosons},\ }\href {https://doi.org/10.1103/PhysRevLett.98.050405} {\bibfield  {journal} {\bibinfo  {journal} {Phys. Rev. Lett.}\ }\textbf {\bibinfo {volume} {98}},\ \bibinfo {pages} {050405} (\bibinfo {year} {2007})}\BibitemShut {NoStop}%
\bibitem [{\citenamefont {Allahverdyan}\ \emph {et~al.}(2004)\citenamefont {Allahverdyan}, \citenamefont {Balian},\ and\ \citenamefont {Nieuwenhuizen}}]{Allahverdyan2004}%
  \BibitemOpen
  \bibfield  {author} {\bibinfo {author} {\bibfnamefont {A.~E.}\ \bibnamefont {Allahverdyan}}, \bibinfo {author} {\bibfnamefont {R.}~\bibnamefont {Balian}},\ and\ \bibinfo {author} {\bibfnamefont {T.~M.}\ \bibnamefont {Nieuwenhuizen}},\ }\bibfield  {title} {\bibinfo {title} {Maximal work extraction from finite quantum systems},\ }\href {https://doi.org/10.1209/epl/i2004-10101-2} {\bibfield  {journal} {\bibinfo  {journal} {EPL}\ }\textbf {\bibinfo {volume} {67}},\ \bibinfo {pages} {565} (\bibinfo {year} {2004})}\BibitemShut {NoStop}%
\bibitem [{\citenamefont {Pusz}\ and\ \citenamefont {Woronowicz}(1978)}]{pusz1978}%
  \BibitemOpen
  \bibfield  {author} {\bibinfo {author} {\bibfnamefont {W.}~\bibnamefont {Pusz}}\ and\ \bibinfo {author} {\bibfnamefont {S.~L.}\ \bibnamefont {Woronowicz}},\ }\bibfield  {title} {\bibinfo {title} {Passive states and {KMS} states for general quantum systems},\ }\href {https://doi.org/10.1007/BF01614224} {\bibfield  {journal} {\bibinfo  {journal} {Commun. Math. Phys.}\ }\textbf {\bibinfo {volume} {58}},\ \bibinfo {pages} {273} (\bibinfo {year} {1978})}\BibitemShut {NoStop}%
\bibitem [{\citenamefont {Alicki}\ and\ \citenamefont {Fannes}(2013)}]{alicki2013}%
  \BibitemOpen
  \bibfield  {author} {\bibinfo {author} {\bibfnamefont {R.}~\bibnamefont {Alicki}}\ and\ \bibinfo {author} {\bibfnamefont {M.}~\bibnamefont {Fannes}},\ }\bibfield  {title} {\bibinfo {title} {Entanglement boost for extractable work from ensembles of quantum batteries},\ }\href {https://doi.org/10.1103/PhysRevE.87.042123} {\bibfield  {journal} {\bibinfo  {journal} {Phys. Rev. E}\ }\textbf {\bibinfo {volume} {87}},\ \bibinfo {pages} {042123} (\bibinfo {year} {2013})}\BibitemShut {NoStop}%
\bibitem [{\citenamefont {D'Alessio}\ \emph {et~al.}(2016{\natexlab{b}})\citenamefont {D'Alessio}, \citenamefont {Kafri}, \citenamefont {Polkovnikov},\ and\ \citenamefont {Rigol}}]{D'Alessio03052016}%
  \BibitemOpen
  \bibfield  {author} {\bibinfo {author} {\bibfnamefont {L.}~\bibnamefont {D'Alessio}}, \bibinfo {author} {\bibfnamefont {Y.}~\bibnamefont {Kafri}}, \bibinfo {author} {\bibfnamefont {A.}~\bibnamefont {Polkovnikov}},\ and\ \bibinfo {author} {\bibfnamefont {M.}~\bibnamefont {Rigol}},\ }\bibfield  {title} {\bibinfo {title} {From quantum chaos and eigenstate thermalization to statistical mechanics and thermodynamics},\ }\href {https://doi.org/10.1080/00018732.2016.1198134} {\bibfield  {journal} {\bibinfo  {journal} {Adv. Phys.}\ }\textbf {\bibinfo {volume} {65}},\ \bibinfo {pages} {239} (\bibinfo {year} {2016}{\natexlab{b}})}\BibitemShut {NoStop}%
\bibitem [{\citenamefont {Cazalilla}\ \emph {et~al.}(2011)\citenamefont {Cazalilla}, \citenamefont {Citro}, \citenamefont {Giamarchi}, \citenamefont {Orignac},\ and\ \citenamefont {Rigol}}]{cazalilla2011}%
  \BibitemOpen
  \bibfield  {author} {\bibinfo {author} {\bibfnamefont {M.~A.}\ \bibnamefont {Cazalilla}}, \bibinfo {author} {\bibfnamefont {R.}~\bibnamefont {Citro}}, \bibinfo {author} {\bibfnamefont {T.}~\bibnamefont {Giamarchi}}, \bibinfo {author} {\bibfnamefont {E.}~\bibnamefont {Orignac}},\ and\ \bibinfo {author} {\bibfnamefont {M.}~\bibnamefont {Rigol}},\ }\bibfield  {title} {\bibinfo {title} {One dimensional bosons: From condensed matter systems to ultracold gases},\ }\href {https://doi.org/10.1103/RevModPhys.83.1405} {\bibfield  {journal} {\bibinfo  {journal} {Rev. Mod. Phys.}\ }\textbf {\bibinfo {volume} {83}},\ \bibinfo {pages} {1405} (\bibinfo {year} {2011})}\BibitemShut {NoStop}%
\bibitem [{\citenamefont {Watson}\ and\ \citenamefont {Kheruntsyan}(2025{\natexlab{a}})}]{WatsonKheruntsyan2025}%
  \BibitemOpen
  \bibfield  {author} {\bibinfo {author} {\bibfnamefont {R.~S.}\ \bibnamefont {Watson}}\ and\ \bibinfo {author} {\bibfnamefont {K.}~\bibnamefont {Kheruntsyan}},\ }\bibfield  {title} {\bibinfo {title} {Quantum many-body thermal machines enabled by atom-atom correlations},\ }\href {https://doi.org/10.21468/SciPostPhys.18.6.190} {\bibfield  {journal} {\bibinfo  {journal} {SciPost Phys.}\ }\textbf {\bibinfo {volume} {18}},\ \bibinfo {pages} {190} (\bibinfo {year} {2025}{\natexlab{a}})}\BibitemShut {NoStop}%
\bibitem [{\citenamefont {Watson}\ and\ \citenamefont {Kheruntsyan}(2025{\natexlab{b}})}]{watson2025}%
  \BibitemOpen
  \bibfield  {author} {\bibinfo {author} {\bibfnamefont {R.~S.}\ \bibnamefont {Watson}}\ and\ \bibinfo {author} {\bibfnamefont {K.~V.}\ \bibnamefont {Kheruntsyan}},\ }\bibfield  {title} {\bibinfo {title} {Universal principles for sudden-quench quantum {Otto} engines},\ }\href {https://doi.org/10.1103/h1mn-th94} {\bibfield  {journal} {\bibinfo  {journal} {Phys. Rev. E}\ }\textbf {\bibinfo {volume} {112}},\ \bibinfo {pages} {034120} (\bibinfo {year} {2025}{\natexlab{b}})}\BibitemShut {NoStop}%
\bibitem [{\citenamefont {Nautiyal}\ \emph {et~al.}(2024)\citenamefont {Nautiyal}, \citenamefont {Watson},\ and\ \citenamefont {Kheruntsyan}}]{nautiyal2024}%
  \BibitemOpen
  \bibfield  {author} {\bibinfo {author} {\bibfnamefont {V.~V.}\ \bibnamefont {Nautiyal}}, \bibinfo {author} {\bibfnamefont {R.~S.}\ \bibnamefont {Watson}},\ and\ \bibinfo {author} {\bibfnamefont {K.~V.}\ \bibnamefont {Kheruntsyan}},\ }\bibfield  {title} {\bibinfo {title} {A finite-time quantum {Otto} engine with tunnel coupled one-dimensional {Bose} gases},\ }\href {https://doi.org/10.1088/1367-2630/ad57e5} {\bibfield  {journal} {\bibinfo  {journal} {New J. Phys.}\ }\textbf {\bibinfo {volume} {26}},\ \bibinfo {pages} {063033} (\bibinfo {year} {2024})}\BibitemShut {NoStop}%
\bibitem [{\citenamefont {Nautiyal}(2025)}]{PhysRevE.111.054133}%
  \BibitemOpen
  \bibfield  {author} {\bibinfo {author} {\bibfnamefont {V.~V.}\ \bibnamefont {Nautiyal}},\ }\bibfield  {title} {\bibinfo {title} {Out-of-equilibrium quantum thermochemical engine with one-dimensional {Bose} gas},\ }\href {https://doi.org/10.1103/PhysRevE.111.054133} {\bibfield  {journal} {\bibinfo  {journal} {Phys. Rev. E}\ }\textbf {\bibinfo {volume} {111}},\ \bibinfo {pages} {054133} (\bibinfo {year} {2025})}\BibitemShut {NoStop}%
\bibitem [{\citenamefont {Jeon}\ \emph {et~al.}(2015)\citenamefont {Jeon}, \citenamefont {Kim},\ and\ \citenamefont {Yi}}]{jeon2015}%
  \BibitemOpen
  \bibfield  {author} {\bibinfo {author} {\bibfnamefont {E.}~\bibnamefont {Jeon}}, \bibinfo {author} {\bibfnamefont {Y.~W.}\ \bibnamefont {Kim}},\ and\ \bibinfo {author} {\bibfnamefont {J.}~\bibnamefont {Yi}},\ }\bibfield  {title} {\bibinfo {title} {Initial ensemble dependence of {Jarzynski} equality in the thermodynamic limit},\ }\href {https://doi.org/10.1088/1751-8113/48/30/305002} {\bibfield  {journal} {\bibinfo  {journal} {J. Phys. A: Math. Theor.}\ }\textbf {\bibinfo {volume} {48}},\ \bibinfo {pages} {305002} (\bibinfo {year} {2015})}\BibitemShut {NoStop}%
\bibitem [{\citenamefont {Jin}\ \emph {et~al.}(2016)\citenamefont {Jin}, \citenamefont {Steinigeweg}, \citenamefont {De~Raedt}, \citenamefont {Michielsen}, \citenamefont {Campisi},\ and\ \citenamefont {Gemmer}}]{PhysRevE.94.012125}%
  \BibitemOpen
  \bibfield  {author} {\bibinfo {author} {\bibfnamefont {F.}~\bibnamefont {Jin}}, \bibinfo {author} {\bibfnamefont {R.}~\bibnamefont {Steinigeweg}}, \bibinfo {author} {\bibfnamefont {H.}~\bibnamefont {De~Raedt}}, \bibinfo {author} {\bibfnamefont {K.}~\bibnamefont {Michielsen}}, \bibinfo {author} {\bibfnamefont {M.}~\bibnamefont {Campisi}},\ and\ \bibinfo {author} {\bibfnamefont {J.}~\bibnamefont {Gemmer}},\ }\bibfield  {title} {\bibinfo {title} {Eigenstate thermalization hypothesis and quantum {Jarzynski} relation for pure initial states},\ }\href {https://doi.org/10.1103/PhysRevE.94.012125} {\bibfield  {journal} {\bibinfo  {journal} {Phys. Rev. E}\ }\textbf {\bibinfo {volume} {94}},\ \bibinfo {pages} {012125} (\bibinfo {year} {2016})}\BibitemShut {NoStop}%
\bibitem [{\citenamefont {Williamson}(2025)}]{PhysRevE.111.L012102}%
  \BibitemOpen
  \bibfield  {author} {\bibinfo {author} {\bibfnamefont {L.~A.}\ \bibnamefont {Williamson}},\ }\bibfield  {title} {\bibinfo {title} {Modified jarzynski equality in a microcanonical ensemble},\ }\href {https://doi.org/10.1103/PhysRevE.111.L012102} {\bibfield  {journal} {\bibinfo  {journal} {Phys. Rev. E}\ }\textbf {\bibinfo {volume} {111}},\ \bibinfo {pages} {L012102} (\bibinfo {year} {2025})}\BibitemShut {NoStop}%
\bibitem [{\citenamefont {Gong}\ and\ \citenamefont {Quan}(2015)}]{gong2015}%
  \BibitemOpen
  \bibfield  {author} {\bibinfo {author} {\bibfnamefont {Z.}~\bibnamefont {Gong}}\ and\ \bibinfo {author} {\bibfnamefont {H.~T.}\ \bibnamefont {Quan}},\ }\bibfield  {title} {\bibinfo {title} {Jarzynski equality, {Crooks} fluctuation theorem, and the fluctuation theorems of heat for arbitrary initial states},\ }\href {https://doi.org/10.1103/PhysRevE.92.012131} {\bibfield  {journal} {\bibinfo  {journal} {Phys. Rev. E}\ }\textbf {\bibinfo {volume} {92}},\ \bibinfo {pages} {012131} (\bibinfo {year} {2015})}\BibitemShut {NoStop}%
\bibitem [{\citenamefont {Hoang}\ \emph {et~al.}(2018)\citenamefont {Hoang}, \citenamefont {Pan}, \citenamefont {Ahn}, \citenamefont {Bang}, \citenamefont {Quan},\ and\ \citenamefont {Li}}]{hoang2018}%
  \BibitemOpen
  \bibfield  {author} {\bibinfo {author} {\bibfnamefont {T.~M.}\ \bibnamefont {Hoang}}, \bibinfo {author} {\bibfnamefont {R.}~\bibnamefont {Pan}}, \bibinfo {author} {\bibfnamefont {J.}~\bibnamefont {Ahn}}, \bibinfo {author} {\bibfnamefont {J.}~\bibnamefont {Bang}}, \bibinfo {author} {\bibfnamefont {H.~T.}\ \bibnamefont {Quan}},\ and\ \bibinfo {author} {\bibfnamefont {T.}~\bibnamefont {Li}},\ }\bibfield  {title} {\bibinfo {title} {Experimental test of the differential fluctuation theorem and a generalized {Jarzynski} equality for arbitrary initial states},\ }\href {https://doi.org/10.1103/PhysRevLett.120.080602} {\bibfield  {journal} {\bibinfo  {journal} {Phys. Rev. Lett.}\ }\textbf {\bibinfo {volume} {120}},\ \bibinfo {pages} {080602} (\bibinfo {year} {2018})}\BibitemShut {NoStop}%
\bibitem [{\citenamefont {Alhambra}\ \emph {et~al.}(2016)\citenamefont {Alhambra}, \citenamefont {Masanes}, \citenamefont {Oppenheim},\ and\ \citenamefont {Perry}}]{alhambra2016}%
  \BibitemOpen
  \bibfield  {author} {\bibinfo {author} {\bibfnamefont {A.~M.}\ \bibnamefont {Alhambra}}, \bibinfo {author} {\bibfnamefont {L.}~\bibnamefont {Masanes}}, \bibinfo {author} {\bibfnamefont {J.}~\bibnamefont {Oppenheim}},\ and\ \bibinfo {author} {\bibfnamefont {C.}~\bibnamefont {Perry}},\ }\bibfield  {title} {\bibinfo {title} {Fluctuating work: From quantum thermodynamical identities to a second law equality},\ }\href {https://doi.org/10.1103/PhysRevX.6.041017} {\bibfield  {journal} {\bibinfo  {journal} {Phys. Rev. X}\ }\textbf {\bibinfo {volume} {6}},\ \bibinfo {pages} {041017} (\bibinfo {year} {2016})}\BibitemShut {NoStop}%
\end{thebibliography}
\end{document}